\documentclass[times,referee]{aastex63}
\usepackage{multirow}

\hypersetup{linkcolor=red,citecolor=blue,filecolor=cyan,urlcolor=magenta}

\received{...}
\revised{...}
\accepted{...}

\submitjournal{AJ}

\shorttitle{Quest for the upcoming periastron passage of WR 125}
\shortauthors{Arora et al.}
\begin{document}

\title{Quest for the upcoming periastron passage of an episodic dust maker and particle accelerating colliding wind binary: WR 125}

\correspondingauthor{Bharti Arora}
\email{bhartiarora612@gmail.com}

\author{Bharti Arora}
\affiliation{Aryabhatta Research Institute of Observational Sciences (ARIES), Nainital$-$263 002, India}
\affiliation{School of Studies in Physics \& Astrophysics, Pt. Ravishankar Shukla University, Raipur$-$492 010, India}

\author[0000-0002-4331-1867]{J. C. Pandey}
\affiliation{Aryabhatta Research Institute of Observational Sciences (ARIES), Nainital$-$263 002, India}

\author{Micha\"{e}l De Becker}
\affiliation{Space Sciences, Technologies and Astrophysics Research (STAR) Institute, University of Li\`ege, Quartier Agora, 19c, All\'ee du 6 A\^out, B5c, B-4000 Sart Tilman, Belgium}

\author{S. B. Pandey}
\affiliation{Aryabhatta Research Institute of Observational Sciences (ARIES), Nainital$-$263 002, India}

\author{Nand K. Chakradhari}
\affiliation{School of Studies in Physics \& Astrophysics, Pt. Ravishankar Shukla University, Raipur$-$492 010, India}

\author{Saurabh Sharma}
\affiliation{Aryabhatta Research Institute of Observational Sciences (ARIES), Nainital$-$263 002, India}

\author{Brijesh Kumar}
\affiliation{Aryabhatta Research Institute of Observational Sciences (ARIES), Nainital$-$263 002, India}

\begin{abstract}

We have carried out a long term infrared and X-ray investigation of the colliding wind binary WR 125 (WC7 + O9III). The source was monitored using \textit{AstroSat}$-$Soft X-ray Telescope and TIRCAM2 mounted at the back of 3.6 m Devasthal Optical Telescope. WR 125 appeared brighter in near infra-red \textit{K}-band during the years 2017-2021 which is attributed to another episode of dust formation similar to the one reported during the likely periastron passage at the beginning of the 1990s. This is further supported by enhanced emission observed in W1 and W2 bands of \textit{WISE} from 2018-19. By combining archival X-ray data sets with our new measurements, long-term variations have been noticed. The source reaches a lower emission state in June 2020 (close to the recent infrared maximum) which could be due to enhanced absorption of X-rays produced in the colliding wind region by the WC stellar wind close to the periastron in an eccentric orbit. The time interval between the previous and latest X-ray low states may indicate an orbital period of 28-29 years, in fair agreement with the recurrence time of episodic dust production. 
We also discuss published radio measurements in the context of a common picture based on a long period binary scenario. These results allow us to draw relevant guidelines for future multi-wavelength observations of WR 125. 

\end{abstract}


\section{Introduction}
Among the category of hot and massive stars, Wolf-Rayet stars (WRs) are the evolved counterparts of O-type objects. They are characterized by strong stellar winds that unravel gradually deeper layers presenting abundances resulting from the nucleosynthesis at work in the stellar core. While some WRs display enhanced nitrogen and helium abundances (WN stars) as expected from the CNO cycle, other present enhanced carbon and depletion in nitrogen while helium is still highly present. In the latter case, one is dealing with evolved objects conventionally referred to as WC stars, \textit{i.e.} WR stars with enhanced carbon (see e.g. \citealt{2007ARA&A..45..177C}). These objects are characterized by strong and dense stellar winds, likely as a result of a significantly enhanced opacity due to  highly ionized iron in the deeper layer of their atmosphere \citep{2005A&A...432..633G}. Several decades ago, it appeared that some of  WC stars displayed a significant infrared excess, attributed to the presence of circumstellar dust \citep{1972A&A....20..333A}. The exact circumstances for the formation of dust in WC star environments are still far from being completely elucidated. Many WCs often undergo variable dust production, some periodic and others random. The class of episodic dust makers (EDM) shows recurrent infrared bursts that fade away with time \citep{1995IAUS..163..335W,2008RMxAC..33...71W}. The standard interpretation framework for EDM is dust formation in the colliding-wind region (CWR) of a binary systems made of a WC star and another massive companion. The episodic nature of the dust emission is reasonably well explained as a result of colliding stellar winds in a wide, massive binary system with highly elliptical orbit \citep{1990MNRAS.243..662W}. 

Let's clarify that the physics of CWRs can also be investigated at other wavelengths. Massive binary systems produce thermal X-ray emission that originates from the interaction region of the winds of the massive binary components, in addition to some contributions from the individual stellar winds \citep{1976SvAL....2..138C,1976SvA....20....2P,1992ApJ...386..265S,2010MNRAS.403.1657P}. The X-ray contribution from the CWR is expected to vary periodically, especially if the stars move in an elliptic orbit. The separation (D) varies so that the CWR moves periodically in and out deeper regions of the stellar winds. In particular, in the case of adiabatic shocks, the X-ray emission from the CWR is expected to vary as 1/D, \textit{i.e.} stronger emission close to periastron \citep{1992ApJ...386..265S}. In addition, depending on the inclination of the system, our line of sight to the CWR passes through varying amounts of stellar wind as the orbit progresses, causing variations in the photoelectric absorption by the stellar wind material. Given the high density of WC star winds, the emission measure of the X-ray emitting plasma is expected to be rather high, such as in the case of WR 140 (e.g. \citealt{1990MNRAS.243..662W}).

Massive binaries are also radio emitters. Beside the steady thermal free-free emission produced by individual stellar winds (e.g. \citealt{1975A&A....39....1P,1975MNRAS.170...41W}), in a binary systems the CWR is likely to produce some non-thermal, synchrotron radiation. The latter component is expected to vary as a function of the orbital phase (especially in eccentric orbits) and is characterized by a negative spectral index $\alpha$, defined as $S_\nu$ $\propto$ $\nu^\alpha$. A requirement is the existence of a population of relativistic electrons, hence the so-called class of Particle-Accelerating Colliding-Wind Binaries (PACWBs, \citealt{2013A&A...558A..28D}). The intrinsic synchrotron emission is expected to peak close to periastron, even though free-free absorption (FFA) will modulate the measured flux densities depending on the orientation of the line-of-sight in the system. The brightness of the synchrotron emission arises downstream of a series of energy conversion processes, including energy injection in particle acceleration, that is fed by the kinetic power of the stellar winds. This criterion is especially in favor of CWBs including a WC component, characterized by the strongest stellar winds (\citealt{2013A&A...558A..28D,2017A&A...600A..47D}). Very good examples are once again WR 140 (\citealt{1990MNRAS.243..662W,2005ApJ...623..447D}), or more recently the extreme Apep system made of two WR stars \citep{2021MNRAS.501.2478M}.

It is thus clear that valuable indicators of binarity can be investigated across the electromagnetic spectrum, especially when a WC-type star is involved. These indicators are notably relevant to establish the occurrence of successive periastron passages, and therefore achieve a first determination of their orbital period. The period determination through standard methods (spectroscopy or astrometry) is highly time consuming, and especially proves difficult for orbital periods of several years or a few decades. Searching for indications of orbital modulations through processes related to the CWR on the basis of archived and new data is therefore quite valuable to characterize these objects. In return, the knowledge of their orbital period is crucial to establish future observation strategies and determine the physical dimensions of these systems, that is required for any modeling purpose.

The present study is dedicated to one of the most studied, still poorly characterized WC system: WR 125. This is a long period WC + O binary, whose orbital parameters are not determined yet. This interesting system already benefited of plenty of observations, and would greatly benefit of a multiwavelength investigation to complete already existing data. The objective of this study is to provide updated constraints on the occurrence of its upcoming periastron passage, and to check how updated time series may help to clarify its orbital period. To this aim, we have observed WR 125 using Soft X-ray Telescope (SXT) onboard \textit{AstroSat} and TIFR Near Infrared Imaging Camera-II (TIRCAM2) mounted at the focal plane of the 3.6 m ARIES Devasthal Optical Telescope (DOT). We have also used archival \textit{Swift} and \textit{XMM$-$Newton} X-ray data. Mid-infrared (MIR) investigations were performed using archival data from Near-Earth Object WISE Reactivation (NEOWISE-R) and \textit{Spitzer} Galactic Legacy  Infrared Mid-Plane Survey Extraordinaire (GLIMPSE) surveys. In addition, we also considered radio measurements published over the past decades in our discussion.

This paper is organized as follows. Section \ref{wr125} describes the target and summarizes the outcome of previous works relevant for our purpose. Section \ref{x-ray} presents the X-ray observations and the processing of data used for the present study. The X-ray light curve and spectral analysis are detailed in Section \ref{lc_spec}. The near infrared (NIR) observations of WR 125 are described in Section \ref{nir}. Finally, Section \ref{disc} includes the discussion of our main results, while conclusions are drawn in Section \ref{conc}.

\section{The colliding-wind binary WR\,125}\label{wr125}
WR 125 (MR 93) is a galactic WR binary system classified as WC7ed+O9 III with period $>$6600 days \citep{2001NewAR..45..135V}. It undergoes mass loss at 6 $\times$ 10$^{-5}$ $M_{\odot}\,yr^{-1}$ with a terminal velocity of 2900 $km\,s^{-1}$ \citep{1992MNRAS.258..461W}. 

Infrared photometry during 1981-91 showed a 10-fold rise in 3-4 $\mu$m flux during 1990-91 \citep{1992MNRAS.258..461W}. This was interpreted in terms of condensation of dust in its wind close to the periastron passage, i.e. when the density of the shocked gas in the CWR is the highest. Further characterization of this IR excess revealed that it started in 1990 and lasted for $\sim$3 years being maximum during 1992-93. In addition, absorption lines were seen in its spectrum supporting its CWB status \citep{1994MNRAS.266..247W}. No recurrence of the 1990-93 dust formation episode was noticed till 2014 in 2MASS, \textit{Spitzer} GLIMPSE and \textit{WISE} surveys according to \citet{2014MNRAS.445.1253W}. Few hints regarding infrared brightening in the beginning of 2018 have been provided by \citet{2019MNRAS.488.1282W}, suggesting an orbital period of about 28.3 yr.

The first X-ray observation of WR 125 was performed by \textit{Einstein} observatory in 1981. It was found to be a bright X-ray source with an ISM corrected X-ray luminosity of 1.4$\times 10^{33}$ $erg s^{-1}$ in 0.2-4.0 keV energy range \citep{1987ApJ...320..283P}. \textit{ROSAT} also made a pointed observation of WR 125 with an  exposure time  of 2141 s where it was only marginally detected with an X-ray luminosity of $\sim$9.1 x $10^{32}$ $erg$ $s^{-1}$ in 0.2-2.4 keV energy band \citep{1995IAUS..163..512P}.  However, no variation in the X-ray emission from four observations in 2016$-$17 was found by \citet{2019MNRAS.484.2229M}. We note that, at higher energies, \citet{2013A&A...555A.102W} searched gamma-ray emission from WR 125 using 24 months \textit{Fermi-LAT} data. Gamma-ray emission was not detected in the analyzed data-set but could only provide upper limits to the high energy photon flux.

In the radio domain, WR 125 is a strong, variable source which showed signatures of both thermal and non-thermal emission with a spectral index varying from -0.5 to +0.7 during its observation from 1982 to 1989 (\citealt{1986ApJ...303..239A,1992MNRAS.258..461W}). The non-thermal radio emission from WR 125 faded between 1985 and 1988, very likely due to a more pronounced FFA as the system approached periastron passage shortly before its dust-formation episode occurring in 1991. \citet{2004AJ....127.2885C} reported on an additional measurement obtained in 2001 but at one wavelength only, therefore, preventing any conclusion on the nature (thermal or non-thermal) of the radio emission at that epoch.

\section{X-ray monitoring of WR 125}\label{x-ray}

X-ray data of WR 125 obtained with \textit{AstroSat} \citep{2014SPIE.9144E..1SS}, \textit{Swift} \citep{2005SSRv..120..165B}, and \textit{XMM$-$Newton} \citep{2001A&A...365L...1J} from 2016 November to 2020 November have been analyzed. Table \ref{tab:tab1} presents a log of these X-ray observations. A total of 21 epochs of X-ray observations  including  observations from \textit{Einstein} and \textit{ROSAT} are used in the present study. The earlier X-ray observations of WR 125 from \textit{Swift} and \textit{XMM$-$Newton}  which were investigated by \citet{2019MNRAS.484.2229M} have also been considered in the present study. The processing of those X-ray data sets has been performed again in order to maintain the homogeneity and application of latest calibration. The data reduction procedure adopted for each X-ray observatory is explained as follows.

\subsection{AstroSat}\label{astrosat}
We have observed WR 125 with  \textit{AstroSat} on two occasions. Both observations were made with a time interval of $\sim$9 months using SXT \citep{2017JApA...38...29S} as the prime instrument in the photon counting (PC) mode. The Level-1 data files were collected from \textit{AstroSat} data archive\footnote{https://astrobrowse.issdc.gov.in/astro\_archive/archive/Home.jsp}. The task \textsc{sxtpipeline} (AS1SXTLevel2, version 1.4b)\footnote{https://www.tifr.res.in/$\sim$astrosat\_sxt/sxtpipeline.html} was used to process the Level-1 SXT data. Any contamination caused by the charged particles during the passage of the instrument through the South Atlantic Anomaly region or occultation by the Earth were filtered out. Only the grade 0$-$12 events (single-quadruple pixel events) were selected during the cleaning process which eliminates effects of any charged particles incident on the CCD. Hence, the calibration of the source events along with the extraction of Level-2 cleaned event files for every single orbit is performed by this pipeline. A Julia based merger tool was used to merge the cleaned event files of the individual orbits to make a single cleaned event file. Finally, this Level-2 merged cleaned event file was used to extract the source spectra using \textsc{xselect} (V2.4d) package.

For the extraction of the source spectrum, a circular region of radius of 13$'$ centered at the co-ordinates R.A. = 19h 28m 15.61s and Dec. = +19$^\circ$ 33$'$ 21.53$''$ for WR 125 was chosen. The selection of the source size has been made based upon the fact that the SXT point spread function (PSF) is large (Half Power Diameter of 11$'$ ) due to the scattering by the mirrors and attitude control of the satellite \citep{2017JApA...38...29S}. Therefore, it has been suggested to carry out the extraction for a radius of 10$'$-14$'$ which further depends upon the brightness of the source. An unusual high count rate was noticed during the initial 6029 sec interval for the first observation (Obs ID: 9000002152). Therefore, this time interval was discarded  while extracting the source products leaving the total effective exposure time of 14814 sec for this data set. However, the other observation ID was free from such instances. For the background estimation, the composite background spectrum (SkyBkg\_comb\_EL3p5\_Cl\_Rd16p0\_v01.pha) obtained after a deep blank sky observation was used. However, the  ancillary response file (ARF) ``sxt\_pc\_excl00\_v04\_20190608.arf"  and the response matrix file (RMF) ``sxt\_pc\_mat\_g0to12.rmf" provided by the instrument team\footnote{https://www.tifr.res.in/$\sim$astrosat\_sxt/dataanalysis.html} were used for the spectral analysis. The total count rate in the source region is found to be  $\sim$1.5 times larger than the background count rate for both the data-sets. Therefore, the count rates obtained for WR 125 have been considered as an upper limit and are mentioned in Table \ref{tab:tab1} for both observation IDs.

\subsection{Swift}\label{swift}
The \textit{Swift} X-Ray Telescope (XRT) observed WR 125 frequently from 2016 November to  2020 November in photon counting mode. These observations spanning over a time period of $\sim$4 years enabled us to investigate the X-ray emission for a long time considering the large anticipated orbital period of this binary system. The data were processed with \textit{Swift} \textsc{xrtpipeline} (v0.13.3) using calibration files released on 2020 July 24. The resulted cleaned and calibrated event files were used for the extraction of image, light curve, and spectrum for every observation. This was done by selecting standard event grades of 0$-$12 in the \textsc{xselect} (v2.4d) package. In order to extract the source products, a circular region at the source position with 30 arcsec radius was chosen. However, background X-ray emission was estimated from an annular region of 69 arcsec inner and 127 arcsec outer radius around the source region. The RMF provided by the \textit{Swift} team (swxpc0to12s6\_20130101v014.rmf) was used. The task \textsc{xrtmkarf} was used to calculate an ARF for each data set individually by considering the associated exposure maps which took care of the bad columns. The background light curves were subtracted from the extracted source light curves using the HEASoft (v6.21) task \textsc{lcmath} by properly scaling the respective extraction areas. Since the total number of counts in each of the \textit{Swift} spectrum were low, we had to perform their spectral fitting using the Cash-statistic. Therefore, the \textit{Swift} spectra were binned to have minimum 1 count per energy bin with \textsc{grppha}. 

\subsection{XMM-Newton}\label{xmm}
\begin{figure*}
\begin{center}
\includegraphics[scale=0.45]{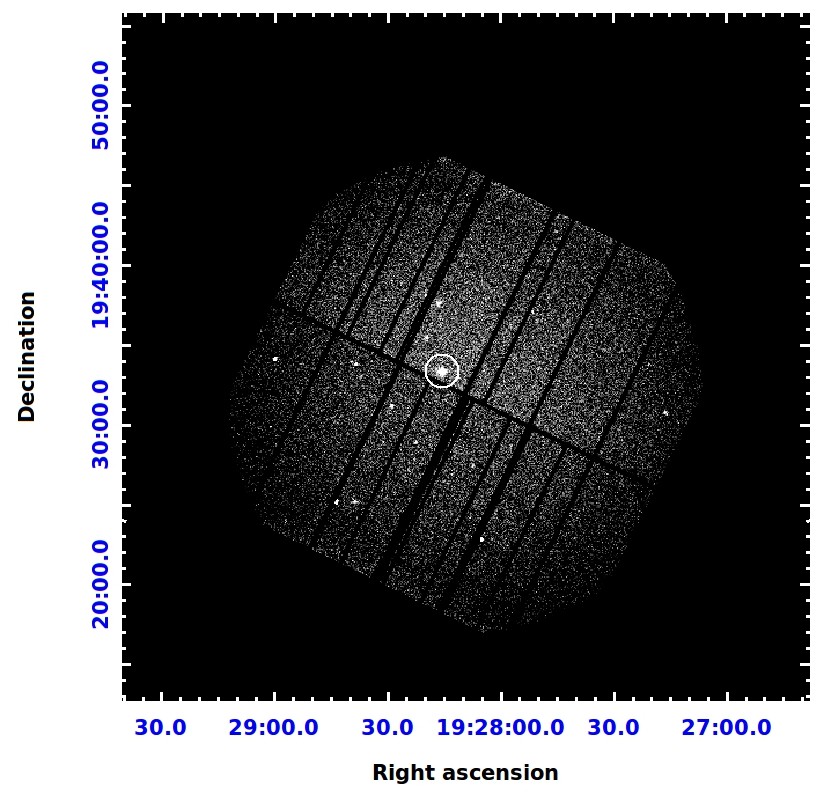}
\caption{\textit{XMM-Newton}$-$PN image from the observation ID 0794581101 in 0.2$-$15.0 keV energy band. The position of WR 125 has been highlighted with a circle.}\label{fig:fig2}
\end{center}
\end{figure*}

WR 125 was observed twice with \textit{XMM$-$Newton}, first in 2017 May and later in 2019 October, with the three European Photon Imaging Camera (EPIC) instruments, \textit{viz.} MOS1, MOS2, and PN. The PN image of WR 125 in 0.2$-$15.0 keV energy range is shown in Figure \ref{fig:fig2}. The data reduction was performed by using the latest calibration files with SAS v17.0.0.

The raw EPIC Observation Data Files (ODF) were pipeline processed using the tasks \textsc{epchain} and \textsc{emchain} for the PN and MOS data, respectively. The SAS task \textsc{evselect} generated the list of event files by considering the good events having pattern 0$-$4 for PN and 0$-$12 for MOS data. The data was found to be unaffected by pile-up after examining with the task \textsc{epatplot}. In order to check whether the data is affected by high background intervals, the light curves were generated considering the single-event (PATTERN = 0) in $>$10 keV energy range for MOS and that in the 10-12 keV band for PN. Selection of the good time intervals was made by removing the high-background intervals where the count rate was higher than 0.35 counts s$^{-1}$ for MOS and  0.4 counts s$^{-1}$ for PN in these light curves.

A circular region of radius of 22 arcsec centered at the source coordinates was selected to extract the EPIC light curves and spectra of WR 125. Background estimation was done from circular region of the same size at source-free regions surrounding the source. The obtained light curves were corrected for good time intervals, dead time, exposure, point-spread function, and background subtraction using the \textsc{epiclccorr} task. The source as well as the background spectra were generated by the task \textsc{evselect}. The dedicated ARF and RMF response matrices required for calibrating the energy and flux axes were calculated by the tasks \textsc{arfgen} and \textsc{rmfgen}, respectively. Back scaling of the extracted spectra was done using the task \textsc{backscale}. In order to have minimum 15 counts per spectral bin, the EPIC spectra were grouped using \textsc{grppha}. Further temporal and spectral analyses were performed using HEASoft version 6.21.

\section{X-ray spectral and light curve analysis}\label{lc_spec}
In order to obtain the best-fit model which explains the X-ray spectrum of WR 125,  we have made use of \textit{XMM-Newton} data as the X-ray spectra from EPIC instruments have better photon statistics than other observatory data. Therefore, we initially decided to jointly fit the MOS1, MOS2, and PN observed spectra of WR 125 in  0.5$-$10.0 keV energy range with the models of the Astrophysical Plasma Emission Code (\textsc{apec}; \citealt{2001ApJ...556L..91S}) as implemented in the X-ray spectral fitting package \textsc{xspec} version 12.9.1 \citep{1996ASPC..101...17A}. Such an optically thin thermal emission model is adequate to reproduce the soft X-ray emission from plasma heated in CWBs (see e.g. \citealt{2015MNRAS.451.1070D,2019MNRAS.487.2624A,2019MNRAS.484.2229M}).

\begin{figure*}
\begin{center}
\includegraphics[scale=0.5]{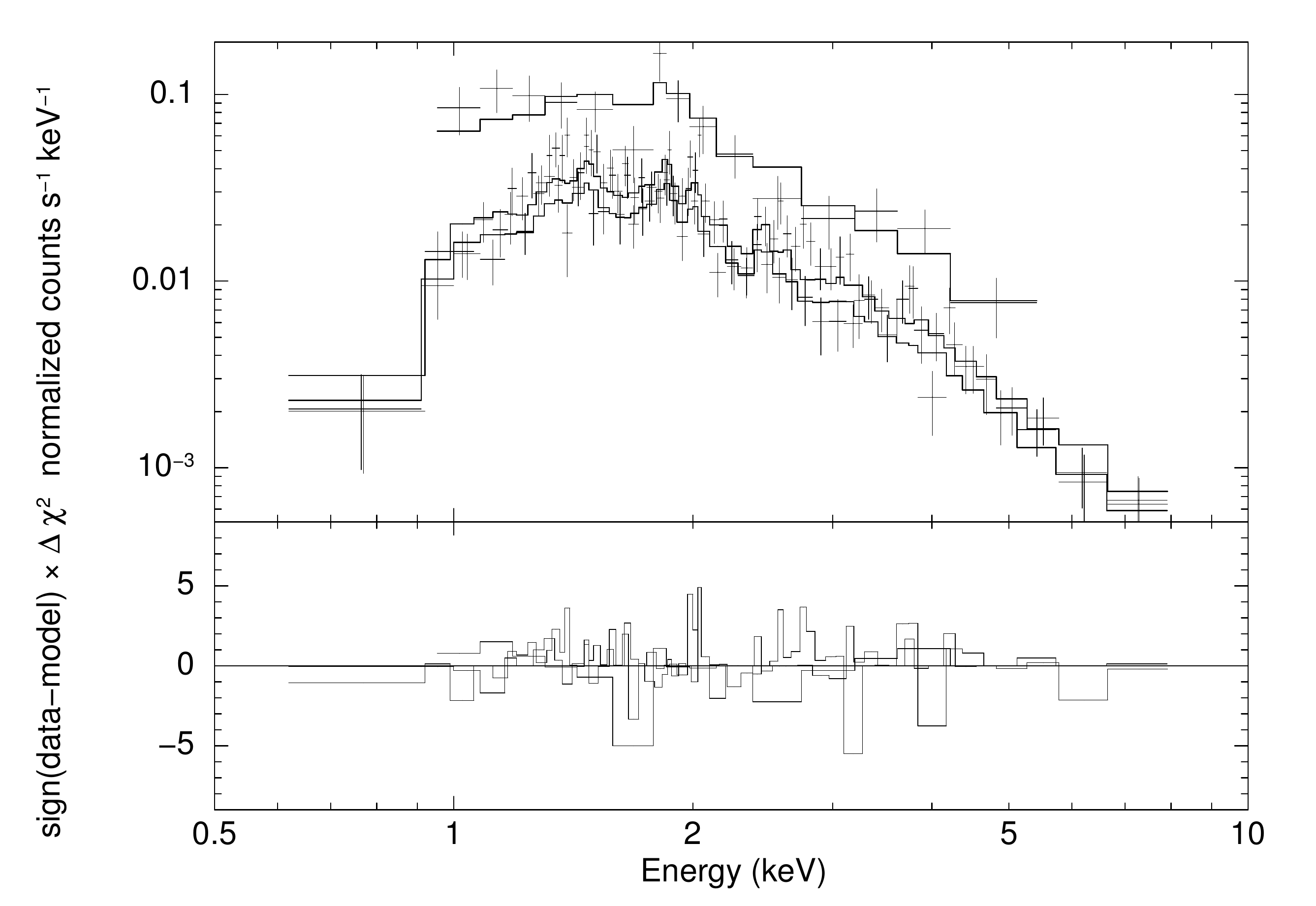}\\
\includegraphics[scale=0.5]{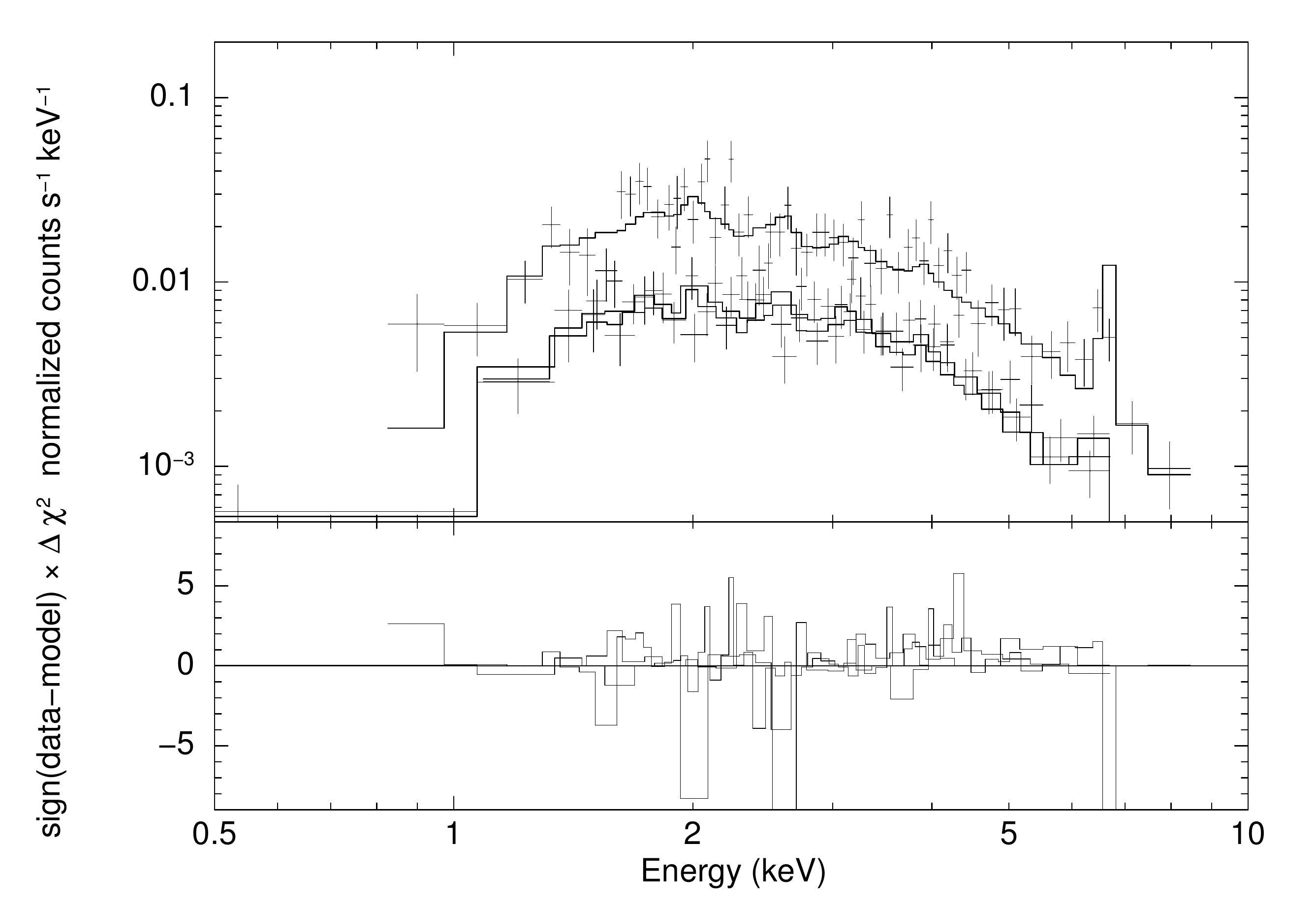}
\caption{\textit{XMM-Newton}$-$EPIC spectra of WR 125 from the observation IDs (a) 0794581101 (\textit{top}) and (b) 0853980101 (\textit{bottom}). The EPIC spectra have been fitted jointly using the model \textsc{phabs*phabs(apec+apec)} with parameters given in Table \ref{tab:tab2} for both the IDs. Both spectra are displayed with the same scale to facilitate their immediate comparison. The lower panels show the residuals in the sense data minus model. \label{fig:fig3}}
\end{center}
\end{figure*}

We first used a two-temperature composite model to account for a simplified distribution of plasma temperature in the system, i.e. \textsc{phabs(ism)*phabs(local)*(apec+apec)}. The interstellar and the circumstellar wind absorption effects were taken into consideration by the model components \textsc{phabs(ism)} and \textsc{phabs(local)}, respectively, with elemental abundances according to \citet{1989GeCoA..53..197A}. The value of N$_{H}^{ISM}$, which is the parameter corresponding to the model component \textsc{phabs(ism)}, was frozen at 0.94$\times$10 $^{22}$ cm$^{-2}$. This value has been derived from the interstellar extinction towards WR 125 using the standard conversion factors \citep{2001NewAR..45..135V,2003A&A...408..581V,2019MNRAS.484.2229M}. The other parameters (local absorption, plasma temperatures and normalization parameters) were kept free in the spectral fitting. The best fit parameters obtained separately for the two observation IDs of \textit{XMM-Newton} are given in Table \ref{tab:tab2} (referred to as Model 1a and Model 1b) while the X-ray spectra of WR 125 with the best-fit model are shown in Figure \ref{fig:fig3}. The low number of counts, especially for the 2019 observation when the X-ray emission is significantly weaker (see the count rates in Table\,\ref{tab:tab1}), leads to unsatisfactory results. The weakening of the X-ray emission between 2017 and 2019 is certainly attributable to a strong increase of the absorption by the stellar winds, despite the expected increase of the intrinsic emission as the system gets closer to periastron. We see that the strong change in soft plasma temperature between 2017 and 2019 is accompanied by a rise of the local absorption column, and a huge difficulty to constrain the emission measure. This is certainly attributable to the lack of required spectral information to constrain the fit in the weaker 2019 spectrum. Let's note that the better constraint (lower relative error) on the normalization parameters of the hotter component comes certainly from the fact the harder part of the spectrum is not significantly affected by the rise of local wind absorption (that suppresses spectral features in the soft part of the spectrum). We clarify that we also tried to use variable abundance models (i.e. \textsc{vapec}), in particular to free the Carbon abundance to account for the nature of the WC wind, but no improvement could be achieved. The relative errors on abundance parameters were much too large to carry any physical meaning. 

The difficulty to achieve satisfactory results using multi-component models led us to use a one-temperature model, i.e. \textsc{phabs(ism)*phabs(local)*apec}, where the interstellar absorption column was frozen to the same value as specified above. The best-fit parameters obtained using this model are given in Table \ref{tab:tab3} and are referred to as Model 2a and 2b for the \textit{XMM-Newton} IDs 0794581101 and 0853980101, respectively. The comparison between Model 2a and Model 2b suggests once again a strong rise of the local absorption in 2019. The lower value for the normalization parameter in 2019 should be considered with caution. As stated above in the framework of the two-temperature model, the lower count number leads inevitably to issues in normalizing the emission component. The much stronger absorption at that epoch removes a significant amount of spectral features that would have been valuable to constrain the emission parameters (i.e. plasma temperature and normalization). In addition, let's keep in mind that considering a unique plasma temperature constitutes a strong approximation in terms of physical description. 

We note that fluxes (observed and corrected for interstellar absorption) are specified in Table \ref{tab:tab2} and in Table \ref{tab:tab3} for the broad (0.5--10.0 keV), the soft (0.5--2.0 keV), and the hard (2.0--10.0 keV) energy bands, respectively for the one-temperature and the two-temperature models. The flux values obtained with the two models (at the same epoch) agree fairly well within error bars. Even though the fits are not of very high quality, they provide a satisfactory basis to determine fluxes in physical units.\\

Since the photon counts are very less in the \textit{Swift} spectra, we decided to fit this by fixing the model parameters to the values as obtained for \textit{XMM-Newton} spectra. The spectrum from each observation ID was fitted individually. For the low count \textit{Swift} spectra, Cash-statistic was used with the spectral binning such that each bin contains at least one count. The \textit{Swift} data obtained closer to the \textit{XMM-Newton} observation ID 0794581101 was fitted with Model 1a. However, the spectra observed in 2019 and later were fitted with Model 1b obtained for \textit{XMM-Newton} ID 0853980101. In spite of this, four \textit{Swift} spectra which could not be fitted well even with the fixed parameters of this model due to very poor net counts. For that, we have made use of the X-ray analysis tool WebPIMMS\footnote{https://heasarc.gsfc.nasa.gov/cgi-bin/Tools/w3pimms/w3pimms.pl} to convert their count rate to flux. Since the \textit{Swift} spectra which could not be fitted well with two temperature plasma components were observed in 2020, therefore, those were fitted with Model 2b.

\begin{figure*}
\begin{center}
\includegraphics[scale=0.45,trim={0.0cm 2.0cm 0.0cm 2.0cm}]{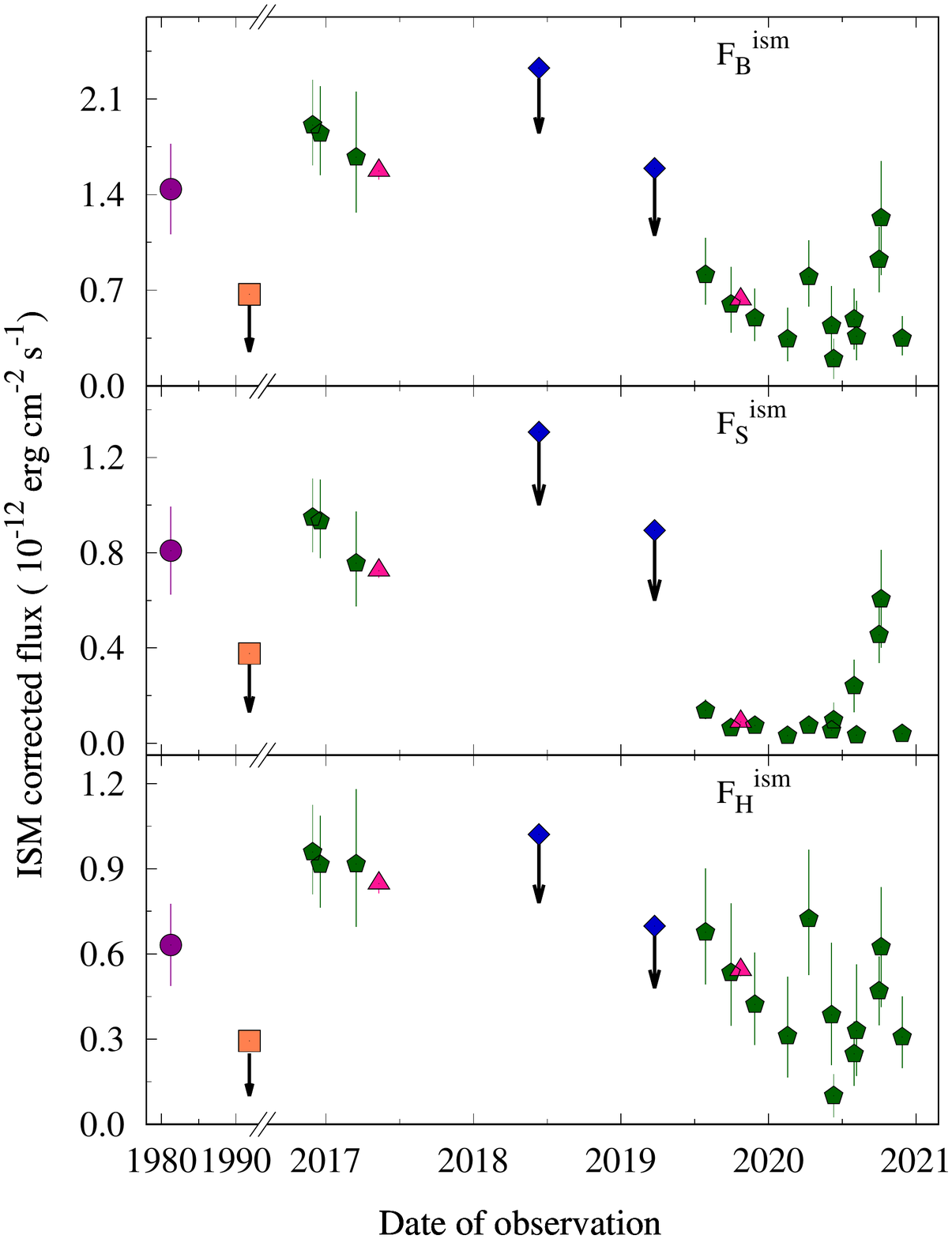}
\caption{Variation of ISM corrected (F$^{ism}$) flux of WR 125. F$_{B}$, F$_{S}$, and F$_{H}$ are the fluxes in the broad (0.5-10.0 keV), soft (0.5-2.0 keV), and hard (2.0-10.0 keV) energy bands, respectively. The purple circle marks the \textit{Einstein}$-$IPC data point, the orange square is the \textit{ROSAT}-PSPC flux, green pentagons are the \textit{Swift}$-$XRT flux measurements while pink triangles correspond to the \textit{XMM-Newton}$-$EPIC observations and blue diamonds show the \textit{AstroSat}$-$SXT flux values.    \label{fig:fig4}}
\end{center}
\end{figure*}

Further, \textit{Einstein} also observed WR 125 with a count rate of 0.0122$\pm$0.0028 counts s$^{-1}$ in the 0.4-4.0 keV energy band on 1981 April 9 using Imaging Proportional Counter (IPC) \citep{2019MNRAS.484.2229M}. However, \textit{ROSAT} only marginally detected WR 125 by making use of Position Sensitive Proportional Counter (PSPC) in 0.1-2.0 keV energy range on 1991 October 28. The upper limit of the WR 125 count rate by \textit{ROSAT}-PSPC is 5.0$\times$10$^{-3}$ counts s$^{-1}$ \citep{2019MNRAS.484.2229M}. Also, the \textit{AstroSat}-SXT spectra of WR 125 were also very poor to be fitted properly in \textsc{xspec}. Therefore, the \textit{Einstein}-IPC, \textit{ROSAT}-PSPC and \textit{AstroSat}-SXT count rate were converted to flux using Model 2a with WebPIMMS only. Finally, the observed X-ray flux ($F^{obs}$) and the ISM corrected X-ray flux ($F^{ism}$) of WR 125 in  broad ($F_{B}$), soft ($F_{S}$), and hard ($F_{H}$) energy bands have been estimated for each of the available spectra. $F^{ism}$ has been plotted with date of observation in Figure \ref{fig:fig4}. Table \ref{tab:tab4} gives the details of the spectral fitting of each of the spectra and also includes the spectral fitting parameters as well as the flux values. 

\begin{figure*}
\begin{center}

\includegraphics[scale=1.0]{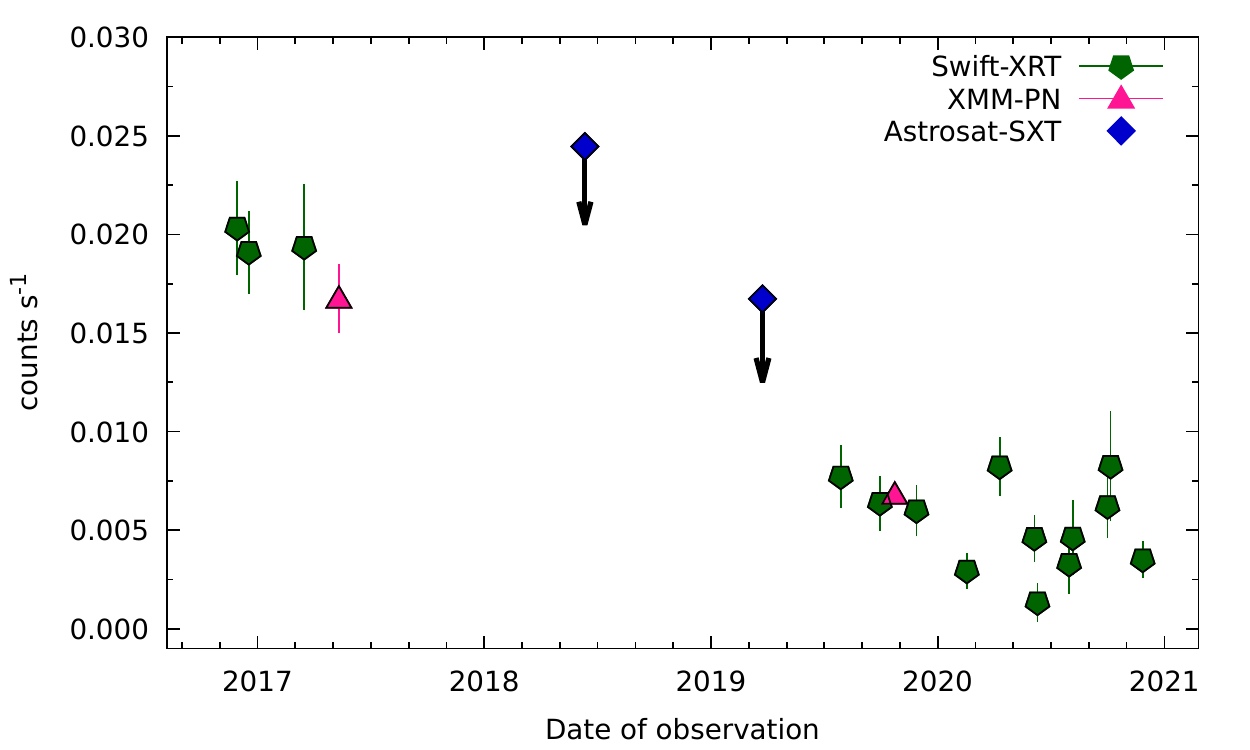}
\caption{X-ray light curve of WR 125 in 0.3$-$10.0 keV energy band as observed by \textit{Swift}$-$XRT (green pentagons), \textit{XMM-Newton}$-$PN (pink triangles), and \textit{AstroSat}$-$SXT (blue diamonds). The PN and SXT count rate has been converted to that of the \textit{Swift}$-$XRT using WebPIMMS.} \label{fig:fig5}
\end{center}
\end{figure*}

The background subtracted X-ray light curve of WR 125 as observed by \textit{AstroSat}-SXT, \textit{Swift}-XRT and \textit{XMM-Newton}-PN in the 0.3-10.0 keV energy band are shown in Figure \ref{fig:fig5}. The average count rate of an individual observation was taken as a single data point in the light curve as the light curve of individual observations was non-variable. Here, the \textit{AstroSat}-SXT count rate in the 0.5-7.0 keV band has been converted to that of \textit{Swift}-XRT in the 0.3-10.0 keV band using WebPIMMS with Model 2a. However, \textit{XMM-Newton}-PN count rate from observation ID 0794581101 and 0853980101 was converted to \textit{Swift}-XRT using Model 2a and 2b, respectively.  As seen in the light curve, the X-ray flux switches from a higher emission state in 2017 to a lower emission state in 2020. The ratio of the maximum to the minimum count rate was found to be 15.33$\pm$11.70.    

\section{NIR monitoring of WR 125}\label{nir}
The NIR imaging of WR 125 was performed on the nights of 2017 October 17, 2020 October 09, and 2021 March 01 using TIRCAM2 (\citealt{2012BASI...40..531N,2018JAI.....750003B}) mounted on the 3.6 m DOT \citep{2018BSRSL..87...29K}. Observations were carried out in J, H, and K filters with central wavelengths at 1.20, 1.65 and 2.19 $\mu$m, respectively. TIRCAM2 has a field of view (FoV) of $\sim$86.5$''$ $\times$ 86.5$''$ with a pixel scale of 0.169$''$. The full width at half maximum (FWHM) of point source PSF was measured to be sub-arcsecond in all JHK bands during the observations.  In addition to the science frames, several dark frames with a similar exposure to the science frame  and sky flat frames were also taken during the observations to account for the  thermal current and the non-uniformities present in the science frames, respectively.
The science frames were taken by dithering the source position to three different locations on the CCD.  For WR 125, eleven frames of 1.0 s exposure were obtained in J band in each of the three dithered positions. However, the same number of frames with 0.5 s exposure time were obtained in both H and K bands.

For the photometric reduction, the Image Reduction and Analysis Facility (IRAF) software\footnote{http://iraf.net/} was used. Similar strategy was followed for all the filters. At first, a master dark frame was generated by average combining all the corresponding dark frames. After subtracting master dark frame from each scientific frame, flat fielding correction was applied by dividing the dark corrected frames by the normalized master flat frame. 
We have constructed the master sky frame in each of the filters by median combining the science frames with dithered source positions of the corresponding NIR band. This was possible because the acquired target field was not crowded. The sky subtracted science frames were finally aligned and combined to yield a good signal-to-noise ratio image of the source. The aperture photometry was performed in all the J, H, and K bands as WR 125 was isolated in all the science frames.  The instrumental magnitudes of WR 125 at each epoch of observations were calibrated by using color equations generated from the instrumental and 2MASS photometry of the other stars in the same field of view (see \citealt{2020MNRAS.498.2309S}). The absolute JHK magnitudes obtained for WR 125 are given in Table \ref{tab:tab5}.




\section{Discussion}\label{disc}

\subsection{Dust emission}\label{dustdisc}
\begin{figure*}
\begin{center}
\includegraphics[scale=1.0]{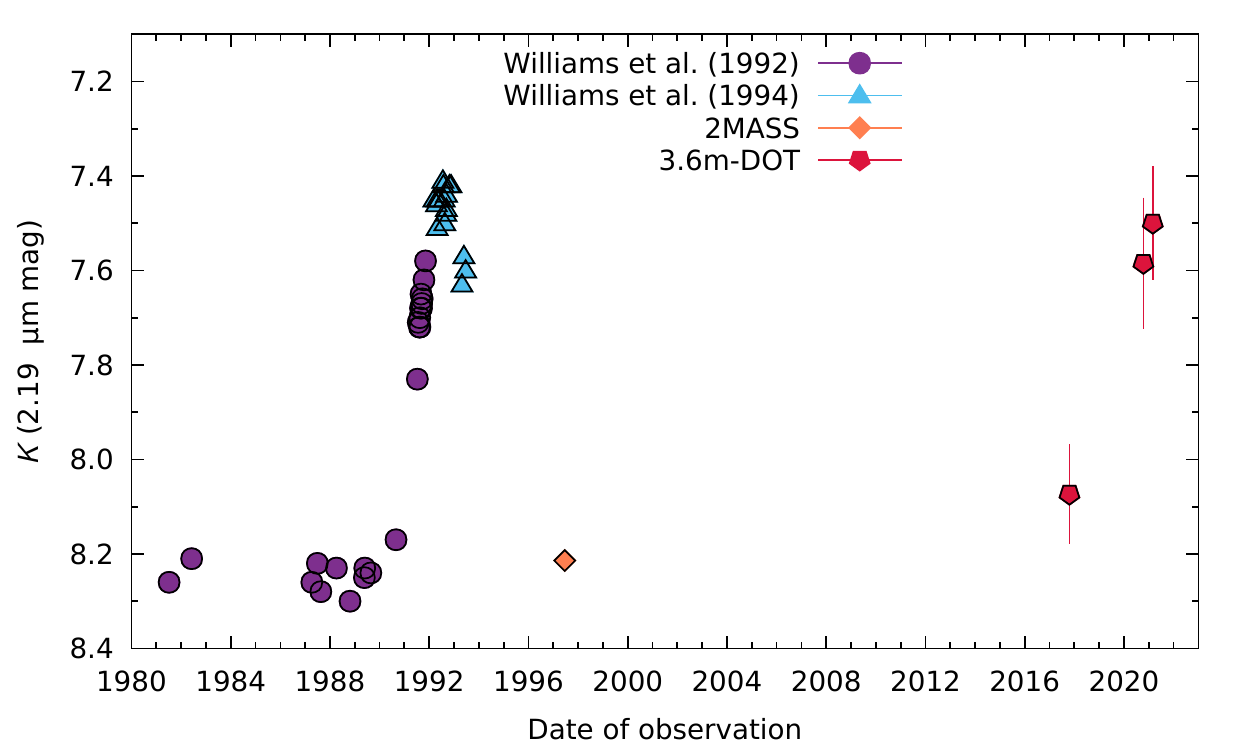}
\caption{\textit{K}-band light curve of WR 125 by combining data from literature, 2MASS survey, and the present study. The accuracy of data from \citet{1992MNRAS.258..461W,1994MNRAS.266..247W} is 0.05 mag.} \label{fig:fig6}
\end{center}
\end{figure*}

The dust emission is efficiently revealed at wavelengths longer than J- and H-bands \citep{1992MNRAS.258..461W,1994MNRAS.266..247W}. Therefore, a comparison of \textit{K}-band magnitude of WR 125 from the present study has been made with the estimations provided in literature as well as in the 2MASS survey in Figure \ref{fig:fig6}. It is clear that during the episode of dust formation in the early 1990's, the \textit{K}-magnitude varied from 8.26 (continuum level) to 7.42 (during outburst) over the years $\sim$1992-1993 \citep{1994MNRAS.266..247W}. Additionally, the \textit{K}-band magnitude obtained from this study shows enhanced emission indicating that the recent measurements were also made during another episode of dust formation. 

\begin{figure*}
\begin{center}
\includegraphics[scale=0.50,trim={0.0cm 2.0cm 0.0cm 5.5cm}]{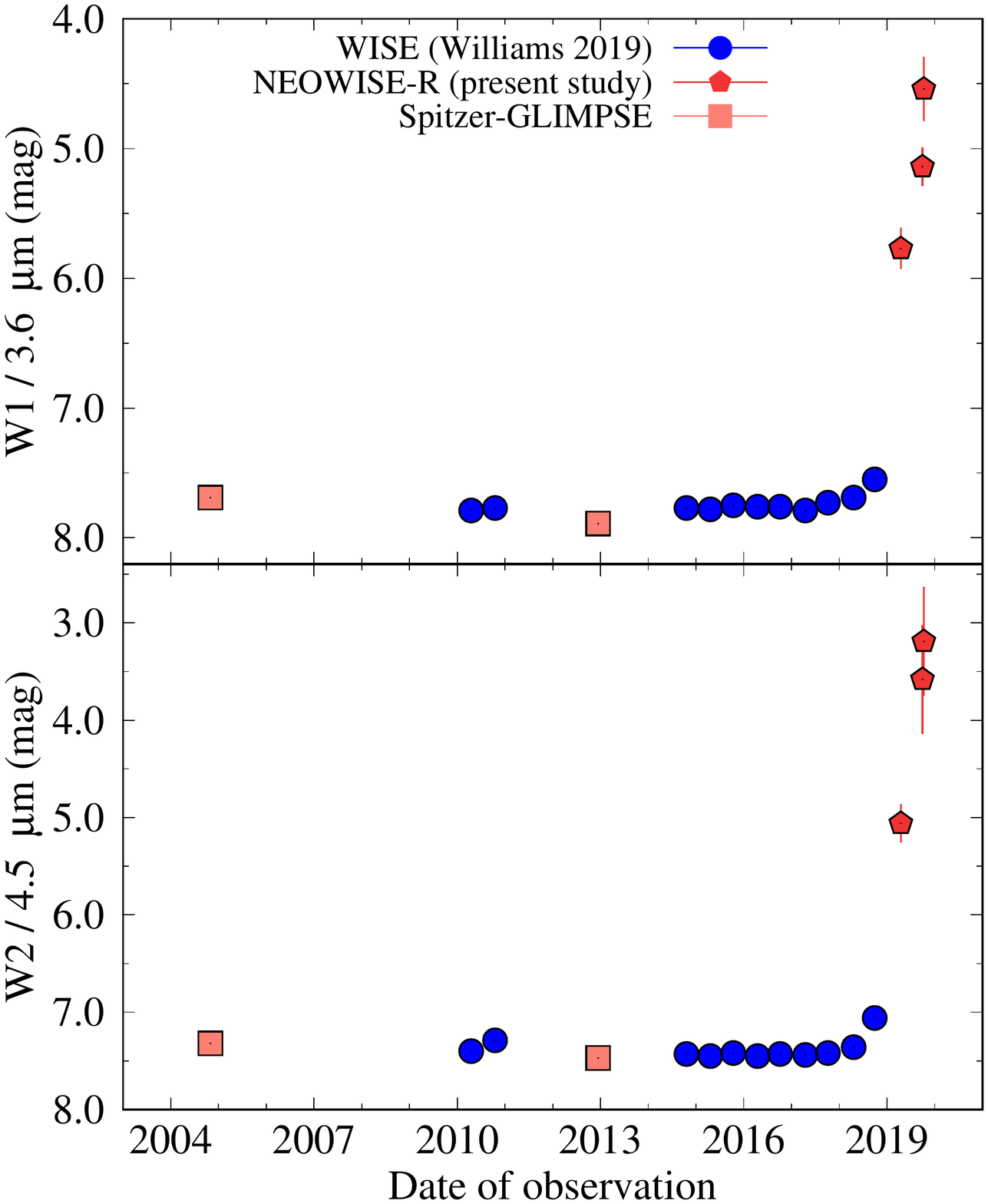}
\caption{Mid infrared light curve of WR 125 by combining \textit{WISE} \textit{W1} and \textit{W2} bands data from \citet{2019MNRAS.488.1282W} (blue filled circles) and the recent NEOWISE-R observations (red pentagons). The \textit{Spitzer}-GLIMPSE magnitudes in the 3.6 and 4.5 $\mu$m wave-bands are also presented (salmon squares). The latest NEOWISE-R data-points have been corrected for the saturation effects as mentioned in \citet{2014ApJ...792...30M}.} \label{fig:fig7}
\end{center}
\end{figure*}

Two time scales deserve to be considered when investigating episodic dust production episodes: one is the time required for the dimming of the IR emission to the pre-outburst level and the other is the time interval between two such consecutive incidences. The dimming of the IR emission is caused by the fading of the dust when being taken away by the stellar wind and is not replenished further. This duration depends upon the terminal velocity of the winds and the amplitude of the dust emission. On the other side, the orbital period of the binary system governs the interval between the two episodes. As the previous dust formation episode was associated to the time of periastron passage of this eccentric CWB system, the enhanced IR emission from the source during our new observations suggests that next periastron passage of WR 125 has been captured. Previous episode of enhanced emission by dust followed by fading lasted for about 3-4 years which is also reflected in the present three epochs of IR observations performed with a time interval of $\sim$3 years. Due to the absence of data in between the latest epochs, the time sampling of this outburst is rather poor. Our assertions are supported by \citet{2019MNRAS.488.1282W} which also gave an indication of IR brightening of WR 125 in the beginning of 2018 using NEOWISE-R observations. We have also retrieved NEOWISE-R \citep{2011ApJ...731...53M} Single Exposure Source observations of WR 125 made in \textit{W1} (3.4 $\mu$m) and \textit{W2} (4.6 $\mu$m) bands in the year 2019 from NASA$/$IPAC Infrared Science Archive (IRSA\footnote{https://irsa.ipac.caltech.edu/}). The NEOWISE-R observations are severely affected by saturation for the sources brighter than \textit{W1} $\cong$ 8 mag and  \textit{W2} $\cong$ 7 mag. Since WR 125 was noticed to be relatively brighter in 2019, therefore, its \textit{W1} and \textit{W2} magnitudes were corrected using offsets from \citet{2014ApJ...792...30M} for 2019 observations. Additionally, the \textit{Spitzer}-GLIMPSE \citep{2003PASP..115..953B,2009PASP..121..213C} observations  in the 3.6 and 4.5 $\mu$m wave-bands further augmented the \textit{W1} and \textit{W2} magnitudes of WR 125, respectively. A comparison of \textit{W1}  and \textit{W2} bands magnitudes given in \citet{2019MNRAS.488.1282W} with those obtained in 2019 and GLIMPSE survey results are shown in Figure \ref{fig:fig7}. The larger error bars associated with the latest NEOWISE-R data-sets display the uncertainty in the available offsets \citep{2014ApJ...792...30M}. Here, the GLIMPSE 3.6 and 4.5 $\mu$m magnitudes were directly compared with those of \textit{W1} and \textit{W2} magnitudes, respectively, without any adjustments for the photometric bands (similar to \citealt{2019MNRAS.488.1282W}). It is clear that the source showed signatures of IR brightening in 2018 which further enhanced in the year 2019.\\ 

Under the reasonable assumption that dust production is enhanced at periastron, the recent rise of the infrared emission very likely indicates WR 125 underwent periastron passage close to 2019-2021. The previous infrared peak occurred in the early 1990's. A time interval of about 28-29 years can thus be considered as a candidate for the orbital period. One has to be careful as a the sole measurement of the two peaks doesn't allow us to rule out an orbit twice as short.

If we consider the first NIR burst started in 1990$-$1994 and peaking during the year 1992.5, then considering the period of 14$-$15 years, the next burst can be expected during the years 2004$-$2008, with a peak around 2006.5. However, the single MIR Spitzer-GLIMPSE observation in year 2004.83 (Figure \ref{fig:fig7}) does not present any significant rise suggesting that a burst was starting to develop at that time. Let's note however that the time span of the burst and the uncertainty on the time of maximum emission (the peak was not fully reached at the time of our most recent measurements) prevent us to rush into any firm conclusion. 

\begin{figure*}
\begin{center}
\includegraphics[scale=1.0]{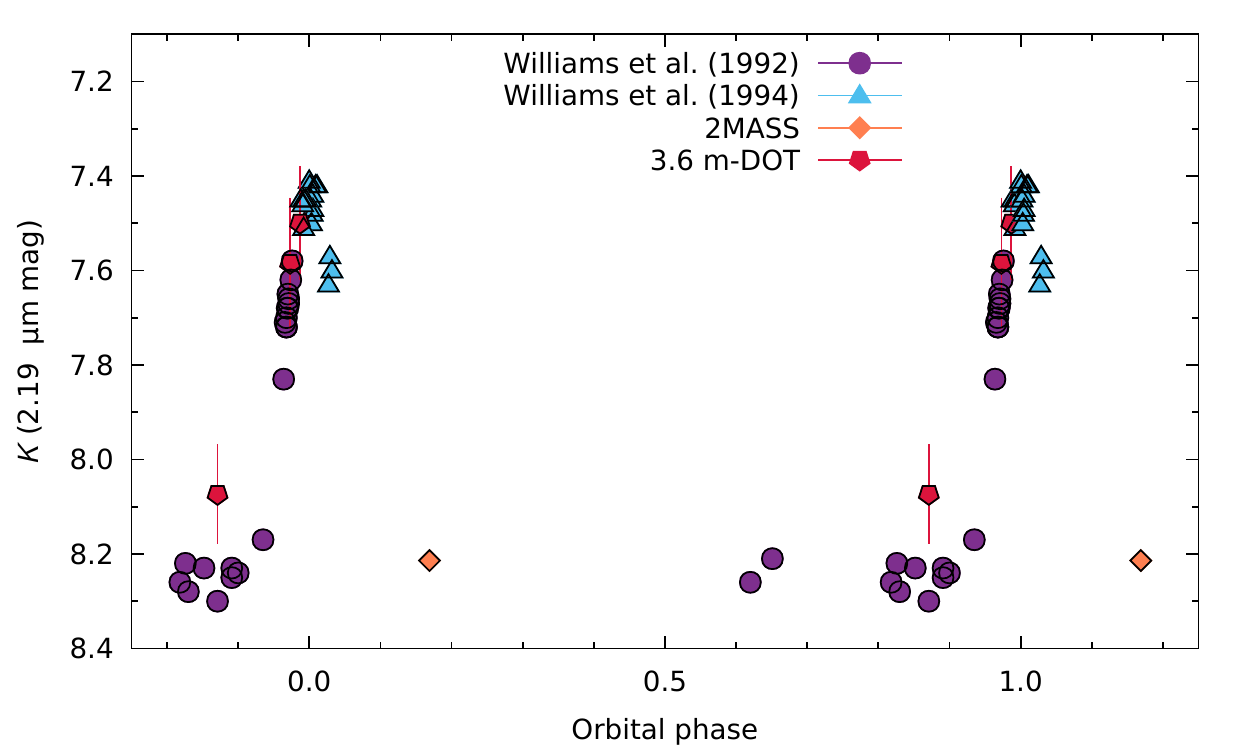}
\caption{\textit{K}-band folded light curve of WR 125 by combining data from literature, 2MASS survey, and the present study. Data was folded over an orbital period of 29 years taking the epoch of maximum brightness as the zero orbital phase.} \label{fig:fig8}
\end{center}
\end{figure*}

To illustrate the longer period possibility, the \textit{K}-band light curve of WR 125 has been folded with an orbital period of 29 years, as shown in Figure \ref{fig:fig8}, taking the epoch of maximum brightness as the zero orbital phase. This plot suggests that new observations are matching well with the previous estimations and supports the hypothesis of 28-29 years orbital period. However, it is worth to mention here that the phase-folded curve with 14$-$15 yrs period is as good as the one with the longer period. Therefore, phase folding of the current time-series data of WR 125 is not enough to constrain its orbital period.
 
In summary,
\begin{enumerate}
\item[-] the 2019-2021 outburst in IR wavelengths suggests the source has undergone another periastron passage recently favoring the dust formation.

\item[-] the time interval between the 1990's and present dust formation episodes indicate an orbital period of 28-29 years. But the possibility of an orbital period twice as short (\textit{i.e.} 14-15 years) emerges here.  

\item[-]  a MIR data point in between the two main peaks does not present any strong rise attributable to a another burst at an intermediate epoch. Even though we caution that a better sampling is needed to achieve a formal confirmation, but this suggests the 14-15 year period is less likely.

\end{enumerate}

\subsection{X-ray emission}\label{Xdisc} 
The main features arising from the long term X-ray monitoring are summarized below:
\begin{enumerate}
\item[-] the observed as well as ISM corrected X-ray flux from WR 125 present long term variations.
\item[-] the X-ray emission switches from some kind of high state in 2016-2017, to a low state in at the end of 2019 and in 2020.
\item[-] the low state is clearly noticed in the soft band, but the hard band is much less conclusive. In addition, best-fit models for \textit{XMM-Newton} data show a stronger absorption column N$^{local}_{H}$  during the low state at the end of 2019, as compared to the high state in 2017.
\item[-] if one relies on the \textit{ROSAT} measurement in 1991 showing a quite low emission state, it suggests a recurrence in the occurrence of the low emission state.
\item[-] a brief rise of the X-ray emission is measured in the second half of 2020, followed by drop of the emission down to the low state plateau. The current time series thus does not reveal any trend leading to a return to a high state as measured in 2016-2017.
\end{enumerate} 

In their recent study, \citet{2019MNRAS.484.2229M} speculate on the occurrence of the minimum emission close to periastron passage. A minimum X-ray emission is tentatively attributed to a more pronounced absorption of X-rays produced in the CWR by the dense and thick WC wind. This interpretation may be compatible with the rather broad low state essentially measured in the soft band, where photoelectric absorption is active. Such an enhanced absorption is expected to happen close to periastron, where the CWR is more deeply surrounded by the WC wind material. We note that the expected dependence of the 1/D trend for adiabatic colliding winds is not revealed in our data set. This would require high quality data over a significant part of the orbit, especially in a part of the orbit that is not that much modulated by a strong absorption. In other words, during the low state, the effect of absorption completely dominates the effect of the reduced separation on the intrinsic X-ray emission in the CWR. Such a situation is similar to what has been observed for WR 140 \citep{1990MNRAS.243..662W}. The start and duration of a low state due to enhanced absorption by the wind depend on the exact geometric configuration of the system (separation, inclination, etc.). As a result, even though the low state is likely related to periastron, the exact time of periastron passage cannot be determined with accuracy, given the available information on the system. We also clarify that in such a long period system, we can certainly reject the hypothesis of an eclipse of the CWR by one of the stars as observed in the short period massive binary WR 121a \citep{2020ApJ...891..104A}. 

Based on the features summarized above, one may speculate on the recurrence of the minimum emission. Given the duration of the recent low state, it is not possible to provide any accurate estimate of the time interval between two minima. The order of magnitude is 28-29 years. In a scenario where these low states are associated to periastron passages, this time interval is a candidate for the orbital period of WR 125. However, we caution that the time sampling of the X-ray emission over such a long period is very sparse. On the basis of the available X-ray data alone, one cannot reject the idea that such a low state event happened just in between, reducing the estimate of the orbital period by a factor 2.

\subsection{Synchrotron radio emission}\label{radiodisc}
\begin{figure*}
\begin{center}
\includegraphics[scale=0.60]{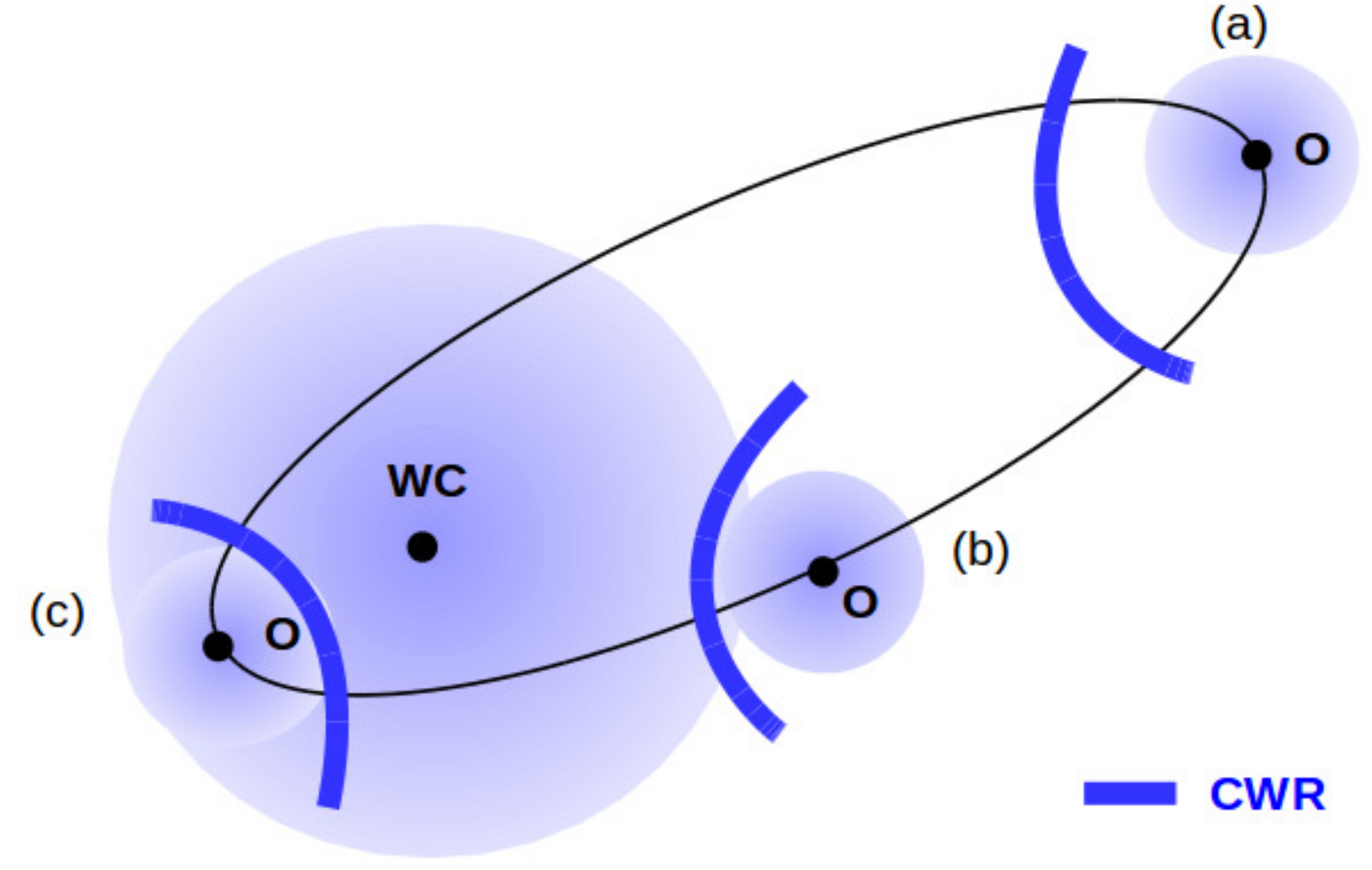}
\caption{Sketch of the likely eccentric orbit of WR 125, illustrating three specific orbital phases: (a) close to apastron, (b) intermediate phase, (c) close to periastron. Blue spheres represent the (not to scale) radio photospheres.} \label{sketchorbit}
\end{center}
\end{figure*}

Only a few measurements have been published, and the radio time series is too sparse to lead to any firm conclusion. The only important result so far is that WR 125 presents significant variations in its behavior, switching between a pure thermal spectrum and a spectrum showing strong indications for non-thermal emission \citep{1986ApJ...303..239A,1992MNRAS.258..461W,2004AJ....127.2885C}. The thermal emission components from the individual stellar winds are steady, and thus do not contribute to the measured variations. In the context of PACWBs, beside the obvious change of physical conditions in the synchrotron emitting region as the separation changes in an eccentric orbit, such a behavior is frequently interpreted in terms of changing FFA. This can happen (i) in an eccentric orbit when the synchrotron emitting region moves in and out the radio photosphere, or (ii) when the CWR is partly hidden behind the thick stellar wind of one of the two stars in the binary system at some orbital phases (line of sight effect). In both cases, this is not the synchrotron radiation process that is inhibited at some epochs, but the disappearance of the non-thermal emission is attributed to FFA. Both aspects are illustrated in Figure\,\ref{sketchorbit}. The blue spheres illustrate the different sizes for radio photospheres of the WC and O-star winds, respectively. Typical sizes for the photosphere can range between tens and thousands of solar radii, depending on the nature of the wind and on the photon frequency (see e.g. \citealt{2019A&A...623A.163D,2020PASA...37...30B}). For a WC wind, the photosphere radius can be of the order of 1000-1500 solar radii at the frequencies of the observations reported by \citet{1986ApJ...303..239A,1992MNRAS.258..461W} and \citet{2004AJ....127.2885C}. Depending on the orbital phase along the eccentric orbit, the CWR can be out of the radio photosphere at apastron, or fully hidden within it at periastron. At intermediate orbital phases, the synchrotron emitting region (coincident with the CWR) may be only partially buried in the photosphere, therefore displaying a spectrum deviating from the pure thermal emission, without any strong non-thermal signature (case (b) in Figure\,\ref{sketchorbit}). However, depending on the orientation of the line of sight, even a not completely buried WCR can be located behind the strong wind of the WC star, and the synchrotron source will thus be significantly absorbed. Let's consider an observer in the orbital plane in Figure\,\ref{sketchorbit}. When observing from the left part, no synchrotron radiation would be measured, but when observing from the bottom some synchrotron radiation could be detected. This orientation effect is more likely to happen not too far from periastron. Otherwise, the alignment between the observer, the WCR and the densest part of the WC wind would not be good enough to lead to a complete absorption of the synchrotron radiation. 

With the above picture in mind, the radio measurements allow us to notice that,
\begin{enumerate}
\item[-] the clear non-thermal radiation identified in 1985 was certainly produced at an orbital phase sufficiently far from periastron.
\item[-] the drop by about a factor 6 of the flux density at 6 cm measured in 1988 (compared to that measured in 1985) is likely pointing to an orbital phase closer to periastron as already pointed out by \cite{1992MNRAS.258..461W}.
\end{enumerate}

\subsection{Toward a more complete picture}\label{completepicture}
We can now summarize the outcomes of Sections\, \ref{dustdisc}, \ref{Xdisc}, and \ref{radiodisc} as 
\begin{enumerate}
\item[-] a strong indication for periastron passage is provided by the dust emission in the infrared. It seems it is occurring close to 2019-2021, in agreement with some complementary elements gathered in X-rays. 
\item[-] a previous periastron passage likely occurred at the beginning of the 1990's. This is indicated by a peak in dust production, and by a drop in the soft X-ray emission at the same epoch. In addition, radio measurements at the end of the 1980's suggest an approaching periastron passage. At the scale of a long period orbit, this is compatible with a likely periastron passage at the beginning of the 1990's. 
\item[-] the orbital period suggested by the infrared and X-ray data sets is about 28-29 years. The existing time series data in these spectral bands do not discard a period of 14-15 years. However, if the constraint on the orbital period published by \citet{2001NewAR..45..135V} is robust (P $>$ 6600 d), this would favor the long period hypothesis.
\end{enumerate}

Our results provide helpful guidelines for future observations, that we can summarize as follows:
\begin{enumerate}
\item[-] a follow-up study of this system in the infrared (and in X-rays) close to year 2034 would be adequate, at it would either be another periastron passage (for the 14-15 year period hypothesis) or a more quiet apastron passage (for 28-29 year period hypothesis). The lack of episodic dust formation episode (and of low X-ray emission state) would point to an apastron passage, in favor of the longest period hypothesis.
\item[-] despite it's high interest for the investigation of non-thermal processes in CWBs, WR 125 was the target of only a few radio observations. It's orbital motion away from periastron would constitute a relevant opportunity to measure the rise of the radio emission (as FFA would decrease), followed by a likely fade out of the emission while approaching apastron (as the separation increases). This would constitute an alternative test to check what is the most realistic orbital period hypothesis.
\item[-] with an orbital period of several years, WR 125 is a valid target for long baseline interferometric measurements aimed at following its orbit and derive orbital parameters. At a distance of about 2 kpc, an orbital period of several years translates into a projected separation of several (tens of) milli-arcseconds, thus easily resolved with current long baseline interferometers operating at the milli-arcsecond scale. In addition, the combination with spectroscopic measurements would allow to derive a three-dimensional orbit (see e.g. \citealt{2017A&A...601A..34L}).
\end{enumerate}

\section{Summary and conclusions}\label{conc}
We have conducted a long term monitoring of the episodic dust maker WR 125 using archive and new data in X-rays and in the (near and mid) infrared. In X-rays, we confirm the switch of the emission to a low state in 2020 (especially in the soft band), that may indicate a recent periastron passage. The drop in the soft X-ray emission could be attributed to a significant photoelectric absorption close to periastron, when the X-ray emission from the CWR may be more quantitatively absorbed by the dense WC wind. The occurrence of a previous low emission state in 1991 suggests a previous periastron passage occurred about 28-29 years before. The infrared emission attributed to dust formation presents a clear rise in 2019-2021, that is confirmed by our very recent measurements. The infrared outburst measured in 2019-2021 is reminiscent of a previous one measured at the beginning of the 1990's, both reasonably coincident with periastron passage in an eccentric long period orbit. Once again, the time interval between these two events is about 28-29 years. We also note that the careful inspection of the radio measurements tend to confirm that the beginning of the 1990's is a very likely epoch of periastron passage. However, the lack of recent radio data prevents any further conclusion to be drawn.

Even though the data analysis reported in this paper seems to confirm that a periastron passage occurred in 2019-2021, one has to be more careful about the orbital period of WR 125. Both the X-ray emission and the dust production suggest a recurrence of about 28-29 years. 
The available X-ray and infrared data do not completely cover the 2004-2005 epoch, where another un-noticed periastron passage could have occurred. In that case, it should have happened right after the unique infrared measurement that exists in 2004 and that shows a low-state magnitude. This reduces the likelihood of such a brightening at that intermediate epoch. Future observations, especially close to year 2034 are needed to definitely lift the slight uncertainty that remains on orbital the period of this system. At that epoch, the system will undergo either a periastron passage (shorter period hypothesis) or an apastron passage (long period hypothesis) and thus strengthens the 28-29 years period.

Finally, we note that our multi-wavelength study of WR 125 allowed us to identify clear guidelines for future observations. This system offers the interesting opportunity to display a rich and diversified physics in its CWR that deserves to be investigated in several spectral domains, in a similar way as the emblematic WC + O system WR 140. In addition, the current knowledge on this system is certainly enough to motivate and organize a long term astrometric (and spectroscopic) campaign aimed at determining it's three-dimensional orbit, that is crucial to interpret and model its behavior across the electromagnetic spectrum.

\acknowledgments
This work has used the data from the Soft X-ray Telescope (SXT) developed at TIFR, Mumbai, and the SXT POC at TIFR is thanked for verifying and releasing the data via the ISSDC data archive and providing the necessary software tools. These results are also based on the observations obtained with \textit{XMM$-$Newton}, an ESA science mission with instruments and contributions directly funded by ESA Member States and NASA. We acknowledge the use of public data from the \textit{Swift} data archive. We also thank the staff at the 3.6 m DOT, Devasthal (ARIES), for their co-operation during TIRCAM2 observations. This publication also makes use of data products from the Near-Earth Object Wide-field Infrared Survey Explorer (NEOWISE), which is a joint project of the Jet Propulsion Laboratory/California Institute of Technology and the University of Arizona. NEOWISE is funded by the National Aeronautics and Space Administration.

%





\bibliographystyle{aasjournal}
\bibliography{ms}{}

\begin{thebibliography}{}
\expandafter\ifx\csname natexlab\endcsname\relax\def\natexlab#1{#1}\fi
\providecommand{\url}[1]{\href{#1}{#1}}
\providecommand{\dodoi}[1]{doi:~\href{http://doi.org/#1}{\nolinkurl{#1}}}
\providecommand{\doeprint}[1]{\href{http://ascl.net/#1}{\nolinkurl{http://ascl.net/#1}}}
\providecommand{\doarXiv}[1]{\href{https://arxiv.org/abs/#1}{\nolinkurl{https://arxiv.org/abs/#1}}}

\bibitem[{{Abbott} {et~al.}(1986){Abbott}, {Beiging}, {Churchwell}, \&
  {Torres}}]{1986ApJ...303..239A}
{Abbott}, D.~C., {Beiging}, J.~H., {Churchwell}, E., \& {Torres}, A.~V. 1986,
  \apj, 303, 239, \dodoi{10.1086/164070}

\bibitem[{{Allen} {et~al.}(1972){Allen}, {Swings}, \&
  {Harvey}}]{1972A&A....20..333A}
{Allen}, D.~A., {Swings}, J.~P., \& {Harvey}, P.~M. 1972, \aap, 20, 333

\bibitem[{{Anders} \& {Grevesse}(1989)}]{1989GeCoA..53..197A}
{Anders}, E., \& {Grevesse}, N. 1989, \gca, 53, 197,
  \dodoi{10.1016/0016-7037(89)90286-X}

\bibitem[{{Arnaud}(1996)}]{1996ASPC..101...17A}
{Arnaud}, K.~A. 1996, in Astronomical Society of the Pacific Conference Series,
  Vol. 101, Astronomical Data Analysis Software and Systems V, ed. G.~H.
  {Jacoby} \& J.~{Barnes}, 17

\bibitem[{{Arora} \& {Pandey}(2020)}]{2020ApJ...891..104A}
{Arora}, B., \& {Pandey}, J.~C. 2020, \apj, 891, 104,
  \dodoi{10.3847/1538-4357/ab7337}

\bibitem[{{Arora} {et~al.}(2019){Arora}, {Pandey}, \& {De
  Becker}}]{2019MNRAS.487.2624A}
{Arora}, B., {Pandey}, J.~C., \& {De Becker}, M. 2019, \mnras, 487, 2624,
  \dodoi{10.1093/mnras/stz1447}

\bibitem[{{Baug} {et~al.}(2018){Baug}, {Ojha}, {Ghosh}, {Sharma}, {Pandey},
  {Kumar}, {Ghosh}, {Ninan}, {Naik}, {D'Costa}, {Poojary}, {Sandimani}, {Shah},
  {Krishna Reddy}, {Pandey}, \& {Chand}}]{2018JAI.....750003B}
{Baug}, T., {Ojha}, D.~K., {Ghosh}, S.~K., {et~al.} 2018, Journal of
  Astronomical Instrumentation, 7, 1850003, \dodoi{10.1142/S2251171718500034}

\bibitem[{{Benaglia} {et~al.}(2020){Benaglia}, {De Becker}, {Ishwara-Chandra},
  {Intema}, \& {Isequilla}}]{2020PASA...37...30B}
{Benaglia}, P., {De Becker}, M., {Ishwara-Chandra}, C.~H., {Intema}, H.~T., \&
  {Isequilla}, N.~L. 2020, \pasa, 37, e030, \dodoi{10.1017/pasa.2020.21}

\bibitem[{{Benjamin} {et~al.}(2003){Benjamin}, {Churchwell}, {Babler}, {Bania},
  {Clemens}, {Cohen}, {Dickey}, {Indebetouw}, {Jackson}, {Kobulnicky},
  {Lazarian}, {Marston}, {Mathis}, {Meade}, {Seager}, {Stolovy}, {Watson},
  {Whitney}, {Wolff}, \& {Wolfire}}]{2003PASP..115..953B}
{Benjamin}, R.~A., {Churchwell}, E., {Babler}, B.~L., {et~al.} 2003, \pasp,
  115, 953, \dodoi{10.1086/376696}

\bibitem[{{Burrows} {et~al.}(2005){Burrows}, {Hill}, {Nousek}, {Kennea},
  {Wells}, {Osborne}, {Abbey}, {Beardmore}, {Mukerjee}, {Short}, {Chincarini},
  {Campana}, {Citterio}, {Moretti}, {Pagani}, {Tagliaferri}, {Giommi},
  {Capalbi}, {Tamburelli}, {Angelini}, {Cusumano}, {Br{\"a}uninger}, {Burkert},
  \& {Hartner}}]{2005SSRv..120..165B}
{Burrows}, D.~N., {Hill}, J.~E., {Nousek}, J.~A., {et~al.} 2005, \ssr, 120,
  165, \dodoi{10.1007/s11214-005-5097-2}

\bibitem[{{Cappa} {et~al.}(2004){Cappa}, {Goss}, \& {van der
  Hucht}}]{2004AJ....127.2885C}
{Cappa}, C., {Goss}, W.~M., \& {van der Hucht}, K.~A. 2004, \aj, 127, 2885,
  \dodoi{10.1086/383286}

\bibitem[{{Cherepashchuk}(1976)}]{1976SvAL....2..138C}
{Cherepashchuk}, A.~M. 1976, Soviet Astronomy Letters, 2, 138

\bibitem[{{Churchwell} {et~al.}(2009){Churchwell}, {Babler}, {Meade},
  {Whitney}, {Benjamin}, {Indebetouw}, {Cyganowski}, {Robitaille}, {Povich},
  {Watson}, \& {Bracker}}]{2009PASP..121..213C}
{Churchwell}, E., {Babler}, B.~L., {Meade}, M.~R., {et~al.} 2009, \pasp, 121,
  213, \dodoi{10.1086/597811}

\bibitem[{{Crowther}(2007)}]{2007ARA&A..45..177C}
{Crowther}, P.~A. 2007, \araa, 45, 177,
  \dodoi{10.1146/annurev.astro.45.051806.110615}

\bibitem[{{De Becker}(2015)}]{2015MNRAS.451.1070D}
{De Becker}, M. 2015, \mnras, 451, 1070, \dodoi{10.1093/mnras/stv1034}

\bibitem[{{De Becker} {et~al.}(2017){De Becker}, {Benaglia}, {Romero}, \&
  {Peri}}]{2017A&A...600A..47D}
{De Becker}, M., {Benaglia}, P., {Romero}, G.~E., \& {Peri}, C.~S. 2017, \aap,
  600, A47, \dodoi{10.1051/0004-6361/201629110}

\bibitem[{{De Becker} {et~al.}(2019){De Becker}, {Isequilla}, \&
  {Benaglia}}]{2019A&A...623A.163D}
{De Becker}, M., {Isequilla}, N.~L., \& {Benaglia}, P. 2019, \aap, 623, A163,
  \dodoi{10.1051/0004-6361/201834493}

\bibitem[{{De Becker} \& {Raucq}(2013)}]{2013A&A...558A..28D}
{De Becker}, M., \& {Raucq}, F. 2013, \aap, 558, A28,
  \dodoi{10.1051/0004-6361/201322074}

\bibitem[{{Dougherty} {et~al.}(2005){Dougherty}, {Beasley}, {Claussen},
  {Zauderer}, \& {Bolingbroke}}]{2005ApJ...623..447D}
{Dougherty}, S.~M., {Beasley}, A.~J., {Claussen}, M.~J., {Zauderer}, B.~A., \&
  {Bolingbroke}, N.~J. 2005, \apj, 623, 447, \dodoi{10.1086/428494}

\bibitem[{{Gr{\"a}fener} \& {Hamann}(2005)}]{2005A&A...432..633G}
{Gr{\"a}fener}, G., \& {Hamann}, W.~R. 2005, \aap, 432, 633,
  \dodoi{10.1051/0004-6361:20041732}

\bibitem[{{Jansen} {et~al.}(2001){Jansen}, {Lumb}, {Altieri}, {Clavel}, {Ehle},
  {Erd}, {Gabriel}, {Guainazzi}, {Gondoin}, {Much}, {Munoz}, {Santos},
  {Schartel}, {Texier}, \& {Vacanti}}]{2001A&A...365L...1J}
{Jansen}, F., {Lumb}, D., {Altieri}, B., {et~al.} 2001, \aap, 365, L1,
  \dodoi{10.1051/0004-6361:20000036}

\bibitem[{{Kumar} {et~al.}(2018){Kumar}, {Omar}, {Maheswar}, {Pandey}, {Sagar},
  {Uddin}, {Sanwal}, {Bangia}, {Kumar}, {Yadav}, {Sahu}, {Pant}, {Reddy},
  {Gupta}, {Chand}, {Pandey}, {Joshi}, {Jaiswar}, {Nanjappa}, {Purushottam},
  {Yadav}, {Sharma}, {Pandey}, {Joshi}, {Joshi}, {Lata}, {Mehdi}, {Misra}, \&
  {Singh}}]{2018BSRSL..87...29K}
{Kumar}, B., {Omar}, A., {Maheswar}, G., {et~al.} 2018, Bulletin de la Societe
  Royale des Sciences de Liege, 87, 29

\bibitem[{{Le Bouquin} {et~al.}(2017){Le Bouquin}, {Sana}, {Gosset}, {De
  Becker}, {Duvert}, {Absil}, {Anthonioz}, {Berger}, {Ertel}, {Grellmann},
  {Guieu}, {Kervella}, {Rabus}, \& {Willson}}]{2017A&A...601A..34L}
{Le Bouquin}, J.~B., {Sana}, H., {Gosset}, E., {et~al.} 2017, \aap, 601, A34,
  \dodoi{10.1051/0004-6361/201629260}

\bibitem[{{Mainzer} {et~al.}(2011){Mainzer}, {Bauer}, {Grav}, {Masiero},
  {Cutri}, {Dailey}, {Eisenhardt}, {McMillan}, {Wright}, {Walker}, {Jedicke},
  {Spahr}, {Tholen}, {Alles}, {Beck}, {Brandenburg}, {Conrow}, {Evans},
  {Fowler}, {Jarrett}, {Marsh}, {Masci}, {McCallon}, {Wheelock}, {Wittman},
  {Wyatt}, {DeBaun}, {Elliott}, {Elsbury}, {Gautier}, {Gomillion}, {Leisawitz},
  {Maleszewski}, {Micheli}, \& {Wilkins}}]{2011ApJ...731...53M}
{Mainzer}, A., {Bauer}, J., {Grav}, T., {et~al.} 2011, \apj, 731, 53,
  \dodoi{10.1088/0004-637X/731/1/53}

\bibitem[{{Mainzer} {et~al.}(2014){Mainzer}, {Bauer}, {Cutri}, {Grav},
  {Masiero}, {Beck}, {Clarkson}, {Conrow}, {Dailey}, {Eisenhardt}, {Fabinsky},
  {Fajardo-Acosta}, {Fowler}, {Gelino}, {Grillmair}, {Heinrichsen}, {Kendall},
  {Kirkpatrick}, {Liu}, {Masci}, {McCallon}, {Nugent}, {Papin}, {Rice},
  {Royer}, {Ryan}, {Sevilla}, {Sonnett}, {Stevenson}, {Thompson}, {Wheelock},
  {Wiemer}, {Wittman}, {Wright}, \& {Yan}}]{2014ApJ...792...30M}
{Mainzer}, A., {Bauer}, J., {Cutri}, R.~M., {et~al.} 2014, \apj, 792, 30,
  \dodoi{10.1088/0004-637X/792/1/30}

\bibitem[{{Marcote} {et~al.}(2021){Marcote}, {Callingham}, {De Becker},
  {Edwards}, {Han}, {Schulz}, {Stevens}, \& {Tuthill}}]{2021MNRAS.501.2478M}
{Marcote}, B., {Callingham}, J.~R., {De Becker}, M., {et~al.} 2021, \mnras,
  501, 2478, \dodoi{10.1093/mnras/staa3863}

\bibitem[{{Midooka} {et~al.}(2019){Midooka}, {Sugawara}, \&
  {Ebisawa}}]{2019MNRAS.484.2229M}
{Midooka}, T., {Sugawara}, Y., \& {Ebisawa}, K. 2019, \mnras, 484, 2229,
  \dodoi{10.1093/mnras/sty3488}

\bibitem[{{Naik} {et~al.}(2012){Naik}, {Ojha}, {Ghosh}, {Poojary}, {Jadhav},
  {Meshram}, {Sandimani}, {Bhagat}, {D'Costa}, {Gharat}, {Bakalkar}, {Ninan},
  \& {Joshi}}]{2012BASI...40..531N}
{Naik}, M.~B., {Ojha}, D.~K., {Ghosh}, S.~K., {et~al.} 2012, Bulletin of the
  Astronomical Society of India, 40, 531.
\newblock \doarXiv{1211.5542}

\bibitem[{{Panagia} \& {Felli}(1975)}]{1975A&A....39....1P}
{Panagia}, N., \& {Felli}, M. 1975, \aap, 39, 1

\bibitem[{{Pittard} \& {Parkin}(2010)}]{2010MNRAS.403.1657P}
{Pittard}, J.~M., \& {Parkin}, E.~R. 2010, \mnras, 403, 1657,
  \dodoi{10.1111/j.1365-2966.2010.15776.x}

\bibitem[{{Pollock}(1987)}]{1987ApJ...320..283P}
{Pollock}, A.~M.~T. 1987, \apj, 320, 283, \dodoi{10.1086/165539}

\bibitem[{{Pollock} {et~al.}(1995){Pollock}, {Haberl}, \&
  {Corcoran}}]{1995IAUS..163..512P}
{Pollock}, A.~M.~T., {Haberl}, F., \& {Corcoran}, M.~F. 1995, in IAU Symposium,
  Vol. 163, Wolf-Rayet Stars: Binaries; Colliding Winds; Evolution, ed. K.~A.
  {van der Hucht} \& P.~M. {Williams}, 512

\bibitem[{{Prilutskii} \& {Usov}(1976)}]{1976SvA....20....2P}
{Prilutskii}, O.~F., \& {Usov}, V.~V. 1976, \sovast, 20, 2

\bibitem[{{Sharma} {et~al.}(2020){Sharma}, {Ghosh}, {Ojha}, {Pandey}, {Sinha},
  {Pandey}, {Ghosh}, {Panwar}, \& {Pandey}}]{2020MNRAS.498.2309S}
{Sharma}, S., {Ghosh}, A., {Ojha}, D.~K., {et~al.} 2020, \mnras, 498, 2309,
  \dodoi{10.1093/mnras/staa2412}

\bibitem[{{Singh} {et~al.}(2014){Singh}, {Tandon}, {Agrawal}, {Antia},
  {Manchanda}, {Yadav}, {Seetha}, {Ramadevi}, {Rao}, {Bhattacharya}, {Paul},
  {Sreekumar}, {Bhattacharyya}, {Stewart}, {Hutchings}, {Annapurni}, {Ghosh},
  {Murthy}, {Pati}, {Rao}, {Stalin}, {Girish}, {Sankarasubramanian},
  {Vadawale}, {Bhalerao}, {Dewangan}, {Dedhia}, {Hingar}, {Katoch}, {Kothare},
  {Mirza}, {Mukerjee}, {Shah}, {Shah}, {Mohan}, {Sangal}, {Nagabhusana},
  {Sriram}, {Malkar}, {Sreekumar}, {Abbey}, {Hansford}, {Beardmore}, {Sharma},
  {Murthy}, {Kulkarni}, {Meena}, {Babu}, \& {Postma}}]{2014SPIE.9144E..1SS}
{Singh}, K.~P., {Tandon}, S.~N., {Agrawal}, P.~C., {et~al.} 2014, in Society of
  Photo-Optical Instrumentation Engineers (SPIE) Conference Series, Vol. 9144,
  Space Telescopes and Instrumentation 2014: Ultraviolet to Gamma Ray, ed.
  T.~{Takahashi}, J.-W.~A. {den Herder}, \& M.~{Bautz}, 91441S,
  \dodoi{10.1117/12.2062667}

\bibitem[{{Singh} {et~al.}(2017){Singh}, {Stewart}, {Westergaard},
  {Bhattacharayya}, {Chandra}, {Chitnis}, {Dewangan}, {Kothare}, {Mirza},
  {Mukerjee}, {Navalkar}, {Shah}, {Abbey}, {Beardmore}, {Kotak}, {Kamble},
  {Vishwakarama}, {Pathare}, {Risbud}, {Koyande}, {Stevenson}, {Bicknell},
  {Crawford}, {Hansford}, {Peters}, {Sykes}, {Agarwal}, {Sebastian},
  {Rajarajan}, {Nagesh}, {Narendra}, {Ramesh}, {Rai}, {Navalgund}, {Sarma},
  {Pandiyan}, {Subbarao}, {Gupta}, {Thakkar}, {Singh}, \&
  {Bajpai}}]{2017JApA...38...29S}
{Singh}, K.~P., {Stewart}, G.~C., {Westergaard}, N.~J., {et~al.} 2017, Journal
  of Astrophysics and Astronomy, 38, 29, \dodoi{10.1007/s12036-017-9448-7}

\bibitem[{{Smith} {et~al.}(2001){Smith}, {Brickhouse}, {Liedahl}, \&
  {Raymond}}]{2001ApJ...556L..91S}
{Smith}, R.~K., {Brickhouse}, N.~S., {Liedahl}, D.~A., \& {Raymond}, J.~C.
  2001, \apj, 556, L91, \dodoi{10.1086/322992}

\bibitem[{{Stevens} {et~al.}(1992){Stevens}, {Blondin}, \&
  {Pollock}}]{1992ApJ...386..265S}
{Stevens}, I.~R., {Blondin}, J.~M., \& {Pollock}, A.~M.~T. 1992, \apj, 386,
  265, \dodoi{10.1086/171013}

\bibitem[{{van der Hucht}(2001)}]{2001NewAR..45..135V}
{van der Hucht}, K.~A. 2001, \nar, 45, 135,
  \dodoi{10.1016/S1387-6473(00)00112-3}

\bibitem[{{Vuong} {et~al.}(2003){Vuong}, {Montmerle}, {Grosso}, {Feigelson},
  {Verstraete}, \& {Ozawa}}]{2003A&A...408..581V}
{Vuong}, M.~H., {Montmerle}, T., {Grosso}, N., {et~al.} 2003, \aap, 408, 581,
  \dodoi{10.1051/0004-6361:20030942}

\bibitem[{{Werner} {et~al.}(2013){Werner}, {Reimer}, {Reimer}, \&
  {Egberts}}]{2013A&A...555A.102W}
{Werner}, M., {Reimer}, O., {Reimer}, A., \& {Egberts}, K. 2013, \aap, 555,
  A102, \dodoi{10.1051/0004-6361/201220502}

\bibitem[{{Williams}(1995)}]{1995IAUS..163..335W}
{Williams}, P.~M. 1995, in Wolf-Rayet Stars: Binaries; Colliding Winds;
  Evolution, ed. K.~A. {van der Hucht} \& P.~M. {Williams}, Vol. 163, 335

\bibitem[{{Williams}(2008)}]{2008RMxAC..33...71W}
{Williams}, P.~M. 2008, in Revista Mexicana de Astronomia y Astrofisica
  Conference Series, Vol.~33, Revista Mexicana de Astronomia y Astrofisica
  Conference Series, 71--76

\bibitem[{{Williams}(2014)}]{2014MNRAS.445.1253W}
{Williams}, P.~M. 2014, \mnras, 445, 1253, \dodoi{10.1093/mnras/stu1779}

\bibitem[{{Williams}(2019)}]{2019MNRAS.488.1282W}
---. 2019, \mnras, 488, 1282, \dodoi{10.1093/mnras/stz1784}

\bibitem[{{Williams} {et~al.}(1992){Williams}, {van der Hucht}, {Bouchet},
  {Spoelstra}, {Eenens}, {Geballe}, {Kidger}, \&
  {Churchwell}}]{1992MNRAS.258..461W}
{Williams}, P.~M., {van der Hucht}, K.~A., {Bouchet}, P., {et~al.} 1992,
  \mnras, 258, 461, \dodoi{10.1093/mnras/258.3.461}

\bibitem[{{Williams} {et~al.}(1994){Williams}, {van der Hucht}, {Kidger},
  {Geballe}, \& {Bouchet}}]{1994MNRAS.266..247W}
{Williams}, P.~M., {van der Hucht}, K.~A., {Kidger}, M.~R., {Geballe}, T.~R.,
  \& {Bouchet}, P. 1994, \mnras, 266, 247, \dodoi{10.1093/mnras/266.1.247}

\bibitem[{{Williams} {et~al.}(1990){Williams}, {van der Hucht}, {Pollock},
  {Florkowski}, {van der Woerd}, \& {Wamsteker}}]{1990MNRAS.243..662W}
{Williams}, P.~M., {van der Hucht}, K.~A., {Pollock}, A.~M.~T., {et~al.} 1990,
  \mnras, 243, 662

\bibitem[{{Wright} \& {Barlow}(1975)}]{1975MNRAS.170...41W}
{Wright}, A.~E., \& {Barlow}, M.~J. 1975, \mnras, 170, 41,
  \dodoi{10.1093/mnras/170.1.41}

\end{thebibliography}



\clearpage

\begin{deluxetable*}{c c c c c c c}
\tablenum{1}
\tablecaption{Log of X-ray observations of WR 125}\label{tab:tab1}
\tablewidth{0pt}
\tablehead{ 
\colhead{Satellite/}& \colhead{Obs. ID}& \colhead{Obs. Date} & \colhead{Start time} & \colhead{Effective} & \colhead{Avg. Count rate} &\colhead{Offset}\\
\colhead{Detector}& \colhead{} & \colhead{} & \colhead{[UT]} & \colhead{ Exp. (s)} & \colhead{(10$^{-2}$ cps)} & \colhead{($'$)}
}
\startdata
\textit{Swift}/XRT      & 00034826001 &  2016-11-28 & 01:50:58 & 4799   & 2.03$\pm$0.24 &   3.595  \\
\textit{Swift}/XRT      & 00034826002 &  2016-12-17 & 13:26:57 & 4747   & 1.91$\pm$0.21 &   3.845  \\
\textit{Swift}/XRT      & 00034826003 &  2017-03-16 & 06:18:57 & 2292   & 1.94$\pm$0.32 &   1.034  \\
\textit{XMM-Newton}/EPIC& 0794581101  &  2017-05-11 & 08:25:51 & 1307        &24.79$\pm$2.60 &   0.004  \\
\textit{AstroSat}/SXT   & 9000002152    &  2018-06-11 & 05:58:14 & 14814      & $<$ 2.28 &   0.00   \\
\textit{AstroSat}/SXT   & 9000002816    &  2019-03-24 & 11:46:34 & 22550     & $<$ 1.56  &  0.00   \\
\textit{Swift}/XRT      & 00034826004 &  2019-07-28 & 05:00:35 & 4377   & 0.77$\pm$0.16 &   4.047  \\
\textit{Swift}/XRT      & 00034826005 &  2019-09-29 & 06:36:35 & 4300   & 0.64$\pm$0.14 &   1.791  \\
\textit{XMM-Newton}/EPIC& 0853980101  &  2019-10-23 & 07:09:02 & 11518 & 9.92$\pm$0.39 &   0.004  \\
\textit{Swift}/XRT      & 00034826007 &  2019-11-27 & 10:43:35 & 4607   & 0.59$\pm$0.13 &   1.193  \\
\textit{Swift}/XRT      & 00034826008 &  2020-02-16 & 01:18:35 & 4332   & 0.29$\pm$0.09 &   1.450  \\
\textit{Swift}/XRT      & 00034826009 &  2020-04-09 & 08:52:36 & 4475   & 0.82$\pm$0.15 &   1.974  \\
\textit{Swift}/XRT      & 00034826010 &  2020-06-04 & 00:08:36 & 3453  & 0.46$\pm$0.12 &   2.461  \\
\textit{Swift}/XRT      & 00034826011 &  2020-06-09 & 08:57:36 & 1448   & 0.13$\pm$0.10 &   1.492  \\
\textit{Swift}/XRT      & 00034826012 &  2020-07-30 & 01:00:18 & 2397   & 0.33$\pm$0.15 &   3.936  \\
\textit{Swift}/XRT      & 00034826013 &  2020-08-05 & 01:45:35 & 2502   & 0.46$\pm$0.19 &   4.076  \\
\textit{Swift}/XRT      & 00034826014 &	 2020-09-30 & 02:37:34 & 3603	& 0.62$\pm$0.16 & 3.032  \\
\textit{Swift}/XRT      & 00034826015 &  2020-10-05 & 11:54:35 & 1161	& 0.82$\pm$0.28 &   3.877  \\
\textit{Swift}/XRT      & 00034826016 &  2020-11-25 & 00:13:34 & 4807	&  0.35$\pm$0.09 &   3.877  \\
\enddata
\tablecomments{(i) The effective exposure and average count rate of \textit{XMM-Newton} is that of the EPIC-pn detector.\\
(ii) The average count rate has been corrected for background contamination.\\
(iii) \textit{AstroSat} count rates are measured in the 0.5$-$7.0 keV energy range. However, both \textit{Swift} and \textit{XMM-Newton} count rates are measured in the 0.3$-$10.0 keV energy band. }
\end{deluxetable*}


\begin{deluxetable}{l c c}
\tablenum{2}
\tablecaption{Best-fit parameters obtained from joint fitting of \textit{XMM-Newton}$-$EPIC spectra of WR 125 using the model \textsc{phabs*phabs(apec+apec)} between 0.5 and 10.0 \,keV. Observed fluxes (\textit{F$^{obs}$}) and fluxes corrected for ISM absorption (\textit{F$^{ism}$}) are also specified for three bands: 0.5--10.0\,keV (B), 0.5--2.0\,keV (S) and 2.0--10.0\,keV (H).}\label{tab:tab2}
\tablewidth{1pt}
\tablehead{\colhead{Parameter}&  \colhead{XMM1$^a$} & \colhead{XMM2$^b$}  \\
\cline{2-3}
           \colhead{}      &  \colhead{(Model 1a)} & \colhead{(Model 1b)}
}
\startdata
N$_{H}^{ism}$       (10$^{22}$ cm$^{-2}$)  & 0.94$^{\dagger}$ &  0.94$^{\dagger}$  \\
N$_{H}^{local}$     (10$^{22}$ cm$^{-2}$)  &  0.68$^{+0.12}_{-0.12}$ &  2.35$^{1.56}_{-0.59}$  \\
\textit{k}T$_{1}$   (\textit{k}eV)         & 0.96$^{+0.11}_{-0.12}$ &  0.11$^{+0.08}_{-0.04}$  \\
\textit{norm}$_{1}$ (cm$^{-5}$)  & 1.22$^{+0.39}_{-0.35}\,\times\,10^{-3}$ &  $<$\,3.0  \\
\textit{k}T$_{2}$   (\textit{k}eV)         & 3.63$^{+3.10}_{-0.79}$ &  2.55$^{+0.50}_{-0.37}$  \\
\textit{norm}$_{2}$ (10$^{-3}$ cm$^{-5}$)  & 0.76$^{+0.22}_{-0.27}\,\times\,10^{-3}$ &  1.09$^{+0.30}_{-0.22}\,\times\,10^{-3}$ \\
$\chi_{\nu}^{2}$    (dof)                  & 0.99 (125) &  1.42 (119)  \\
\textit{F$_{B}^{obs}$} (10$^{12}$ cm$^{-2}$ s$^{-1}$) & 0.95$^{+0.05}_{-0.11}$ & 0.53$^{+0.02}_{-0.05}$ \\
\textit{F$_{S}^{obs}$} (10$^{12}$ cm$^{-2}$ s$^{-1}$) & 0.21$^{+0.01}_{-0.01}$ & 0.03$^{+0.01}_{-0.01}$ \\
\textit{F$_{H}^{obs}$} (10$^{12}$ cm$^{-2}$ s$^{-1}$) & 0.74$^{+0.05}_{-0.13}$ & 0.50$^{+0.02}_{-0.04}$ \\
\textit{F$_{B}^{ism}$} (10$^{12}$ cm$^{-2}$ s$^{-1}$) & 1.55$^{+0.12}_{-0.07}$ & 0.64$^{+0.07}_{-0.04}$ \\
\textit{F$_{S}^{ism}$} (10$^{12}$ cm$^{-2}$ s$^{-1}$) & 0.74$^{+0.07}_{-0.07}$ & 0.09$^{+0.02}_{-0.01}$ \\
\textit{F$_{H}^{ism}$} (10$^{12}$ cm$^{-2}$ s$^{-1}$) & 0.84$^{+0.07}_{-0.07}$ & 0.55$^{+0.03}_{-0.03}$ \\
\enddata
\tablecomments{(a) XMM1: ID$-$0794581101     (b) XMM2: ID$-$0853980101\\
$^{\dagger}$Reference: \citet{2019MNRAS.484.2229M} \\ 
Here, N$_{H}^{ism}$ and N$_{H}^{local}$ refer to the galactic and circumstellar H-column density. \textit{k}T$_{1}$ and \textit{k}T$_{2}$ represent the temperatures of the two components of the plasma model. $\chi_{\nu}^{2}$ is the reduced $\chi^2$  and  \textit{dof} is the number of degrees of freedom. Errors quoted on different parameters refer to the 90\% confidence level.}
\end{deluxetable}

\begin{deluxetable}{l c c}
\tablenum{3}
\tablecaption{Best-fit parameters obtained from joint fitting of \textit{XMM-Newton}$-$EPIC spectra of WR 125 using the model \textsc{phabs*phabs*apec} between 0.5 and 10.0 \,keV.}\label{tab:tab3}
\tablewidth{0pt}
\tablehead{\colhead{Parameter}&  \colhead{XMM1$^a$} & \colhead{XMM2$^b$}  \\
\cline{2-3}
      \colhead{}   &  \colhead{(Model 2a)} & \colhead{(Model 2b)} 
}
\startdata
N$_{H}^{ism}$       (10$^{22}$ cm$^{-2}$)  & 0.94$^{\dagger}$ &  0.94$^{\dagger}$  \\
N$_{H}^{local}$     (10$^{22}$ cm$^{-2}$)  &  0.20$^{+0.14}_{-0.16}$ &  1.40$^{+0.29}_{-0.32}$  \\
\textit{k}T      (\textit{k}eV)           & 2.47$^{+0.47}_{-0.18}$ &  3.26$^{+0.82}_{-0.41}$  \\
\textit{norm}    (cm$^{-5}$)    & 1.30$^{+0.15}_{-0.16}\,\times\,10^{-3}$ &  0.78$^{+0.12}_{-0.13}\,\times\,10^{-3}$  \\
$\chi_{\nu}^{2}$ (dof)                    & 1.36 (127) &  1.52 (121) \\
\textit{F$_{B}^{obs}$} (10$^{12}$ cm$^{-2}$ s$^{-1}$) & 0.92$^{+0.06}_{-0.05}$ & 0.55$^{+0.02}_{-0.04}$ \\
\textit{F$_{S}^{obs}$} (10$^{12}$ cm$^{-2}$ s$^{-1}$) & 0.19$^{+0.01}_{-0.01}$ & 0.03$^{+0.01}_{-0.01}$ \\
\textit{F$_{H}^{obs}$} (10$^{12}$ cm$^{-2}$ s$^{-1}$) & 0.73$^{+0.04}_{-0.05}$ & 0.51$^{+0.03}_{-0.04}$ \\
\textit{F$_{B}^{ism}$} (10$^{12}$ cm$^{-2}$ s$^{-1}$) & 1.49$^{+0.05}_{-0.10}$ & 0.64$^{+0.04}_{-0.03}$ \\
\textit{F$_{S}^{ism}$} (10$^{12}$ cm$^{-2}$ s$^{-1}$) & 0.67$^{+0.14}_{-0.08}$ & 0.08$^{+0.02}_{-0.01}$ \\
\textit{F$_{H}^{ism}$} (10$^{12}$ cm$^{-2}$ s$^{-1}$) & 0.83$^{+0.06}_{-0.06}$ & 0.56$^{+0.04}_{-0.03}$ \\
\enddata
\tablecomments{(a) XMM1: ID$-$0794581101    (b) XMM2: ID$-$0853980101\\}
\end{deluxetable}

\begin{longrotatetable}
\begin{deluxetable*}{l c c c c c c c c c c c}
\tablenum{4}
\tablecaption{Best-fit parameters obtained from \textit{AstroSat}, \textit{Einstein}, \textit{Rosat}, and \textit{Swift} spectral fitting of WR 125 using different models.}\label{tab:tab4}
\tablewidth{0pt}
\tablehead{
\colhead{Satellite$/$}&  \colhead{Obs. ID$/$} & \colhead{Fitting}  & \colhead{$norm_1$} & \colhead{$norm_2$} & \colhead{$\chi^{2}_{\nu} (dof)$} & \colhead{$F^{obs}_{B}$} & \colhead{$F^{obs}_{S}$} & \colhead{$F^{obs}_{H}$} & \colhead{$F^{ism}_{B}$} & \colhead{$F^{ism}_{S}$} & \colhead{$F^{ism}_{H}$} \\
\cline{4-5}    \cline{7-12}
\colhead{Detector}& \colhead{Sequence No.} & \colhead{Model$^\ddagger$} &  \multicolumn2c{(\rm $10^{-3}$ cm$^{-5}$)}  &\colhead{} & \multicolumn6c{(\rm $10^{-12}$ erg cm$^{-2}$ s$^{-1}$)}  
}
\startdata
\textit{Einstein}$/$IPC & 8680$^\dagger$ & 2a  & - & - & - & 0.69$^{+0.16}_{-0.16}$  & 0.15$^{+0.04}_{-0.04}$  & 0.54$^{+0.12}_{-0.12}$  & 1.44$^{+0.33}_{-0.33}$ & 0.81$^{+0.19}_{-0.19}$ & 0.63$^{+0.15}_{-0.15}$ \\
\textit{ROSAT}$/$PSPC   & RP170260N00$^\dagger$ & 2a  & - & - & - & $<$ 0.32  & $<$ 0.07  & $<$ 0.25 & $<$ 0.67 & $<$ 0.38  & $<$ 0.29 \\
\textit{Swift}$/$XRT      & 00034826001 & 1a &1.58$^{+0.59}_{-0.55}$     &0.82$^{+0.41}_{-0.35}$  & 0.90 ( 85)   & 1.11$^{+0.19}_{-0.17}$   & 0.27$^{+0.04}_{-0.05}$ & 0.85$^{+0.15}_{-0.13}$ & 1.91$^{+0.33}_{-0.29}$ & 0.95$^{+0.16}_{-0.15}$ & 0.96$^{+ 0.17}_{- 0.15}$ \\
\textit{Swift}$/$XRT      & 00034826002 & 1a &1.57$^{+0.73}_{-0.68}$     &0.77$^{+0.56}_{-0.49}$  & 0.84 ( 73)   & 1.07$^{+0.20}_{-0.18}$   & 0.26$^{+0.04}_{-0.05}$ & 0.81$^{+0.15}_{-0.13}$ & 1.85$^{+0.35}_{-0.31}$ & 0.93$^{+0.17}_{-0.16}$ & 0.92$^{+ 0.17}_{- 0.15}$ \\ 
\textit{Swift}$/$XRT      & 00034826003 & 1a &1.16$^{+0.91}_{-0.82}$     &0.85$^{+0.73}_{-0.58}$  & 0.91 ( 34)   & 1.04$^{+0.29}_{-0.25}$   & 0.22$^{+0.05}_{-0.06}$ & 0.82$^{+0.23}_{-0.19}$ & 1.67$^{+0.48}_{-0.40}$ & 0.76$^{+0.22}_{-0.18}$ & 0.92$^{+ 0.26}_{- 0.22}$ \\
\textit{Astrosat}$/$SXT   & 9000002152$^\dagger$ & 2a  & - & - & - & $<$ 1.12 & $<$ 0.25  & $<$ 0.88  & $<$ 2.33  & $<$ 1.30  & $<$ 1.02  \\
\textit{Astrosat}$/$SXT   & 9000002816$^\dagger$ & 2a  & - & - & - & $<$ 0.77 & $<$ 0.17  & $<$ 0.59  & $<$ 1.59  & $<$ 0.89  & $<$ 0.69  \\
\textit{Swift}$/$XRT      & 00034826004 & 1b &$<$ 14.92$\times10^{3}$& 1.31$^{+ 0.46}_{-0.38}$  & 1.04 ( 25)   & 0.67$^{+0.22}_{-0.18}$  & 0.05$^{+0.01}_{-0.02}$  &  0.62$^{+0.20}_{-0.17}$ & 0.81$^{+0.27}_{-0.22}$ & 0.14$^{+0.05}_{-0.04}$  & 0.68$^{+0.22}_{-0.18}$ \\
\textit{Swift}$/$XRT      & 00034826005 & 1b &$<$  5.18$\times10^{3}$& 1.04$^{+ 0.49}_{-0.38}$  & 0.80 ( 15)   & 0.52$^{+0.24}_{-0.18}$  & 0.03$^{+0.01}_{-0.01}$  &  0.49$^{+0.22}_{-0.17}$ & 0.59$^{+0.27}_{-0.21}$ & 0.06$^{+0.03}_{-0.02}$  & 0.53$^{+0.24}_{-0.19}$ \\
\textit{Swift}$/$XRT      & 00034826007 & 1b &$<$  6.61$\times10^{3}$& 0.82$^{+ 0.39}_{-0.30}$  & 0.23 ( 17)   & 0.41$^{+0.18}_{-0.14}$  & 0.03$^{+0.01}_{-0.01}$  &  0.38$^{+0.17}_{-0.13}$ & 0.49$^{+0.22}_{-0.17}$ & 0.07$^{+0.03}_{-0.03}$  & 0.42$^{+0.18}_{-0.14}$ \\
\textit{Swift}$/$XRT      & 00034826008 & 1b &$<$  5.36$\times10^{3}$& 0.62$^{+ 0.39}_{-0.30}$  & 0.68 ( 07)   & 0.29$^{+0.20}_{-0.14}$  & 0.02$^{+0.01}_{-0.01}$  &  0.28$^{+0.19}_{-0.13}$ & 0.34$^{+0.23}_{-0.16}$ & 0.03$^{+0.02}_{-0.01}$  & 0.31$^{+0.21}_{-0.15}$ \\
\textit{Swift}$/$XRT      & 00034826009 & 1b &$<$  2.03$\times10^{3}$& 1.41$^{+ 0.47}_{-0.39}$  & 1.85 ( 23)   & 0.69$^{+0.23}_{-0.19}$  & 0.04$^{+0.01}_{-0.01}$  &  0.66$^{+0.22}_{-0.18}$ & 0.79$^{+0.27}_{-0.22}$ & 0.07$^{+0.02}_{-0.02}$  & 0.72$^{+0.24}_{-0.19}$ \\
\textit{Swift}$/$XRT      & 00034826010 & 1b &$<$  6.37$\times10^{3}$& 0.75$^{+ 0.55}_{-0.37}$  & 0.59 ( 07)   & 0.37$^{+0.25}_{-0.17}$  & 0.03$^{+0.01}_{-0.02}$  &  0.35$^{+0.23}_{-0.16}$ & 0.44$^{+0.29}_{-0.20}$ & 0.05$^{+0.04}_{-0.03}$  & 0.39$^{+0.25}_{-0.18}$ \\
\textit{Swift}$/$XRT      & 00034826011$^\dagger$ & 2b & - & - & - &  0.08$^{+0.06}_{-0.06}$ &  0.007$^{+ 0.005}_{-0.005}$  & 0.08$^{+ 0.06}_{- 0.06}$ & 0.19$^{+0.15}_{-0.15}$ & 0.09$^{+0.07}_{-0.07}$ & 0.10$^{+0.08}_{-0.08}$ \\
\textit{Swift}$/$XRT      & 00034826012$^\dagger$ & 2b & - & - & - &  0.21$^{+0.09}_{-0.09}$ &  0.02$^{+ 0.01}_{-0.01}$  & 0.19$^{+ 0.09}_{- 0.09}$ & 0.49$^{+0.22}_{-0.22}$ & 0.24$^{+0.11}_{-0.11}$ & 0.25$^{+0.11}_{-0.11}$ \\
\textit{Swift}$/$XRT      & 00034826013 & 1b &$<$  8.76$\times10^{3}$& 0.64$^{+ 0.45}_{-0.33}$  & 1.49 ( 06)   & 0.32$^{+0.23}_{-0.15}$  & 0.02$^{+0.01}_{-0.01}$  &  0.30$^{+0.21}_{-0.15}$ & 0.36$^{+0.26}_{-0.17}$ & 0.03$^{+0.02}_{-0.02}$  & 0.33$^{+0.23}_{-0.16}$ \\
\textit{Swift}$/$XRT      & 00034826014$^\dagger$ & 2b & - & - & - &  0.39$^{+0.10}_{-0.10}$ &  0.03$^{+ 0.01}_{-0.01}$  & 0.36$^{+ 0.09}_{- 0.09}$ & 0.92$^{+0.24}_{-0.24}$ & 0.46$^{+0.12}_{-0.12}$ & 0.47$^{+0.12}_{-0.12}$ \\
\textit{Swift}$/$XRT      & 00034826015$^\dagger$ & 2b & - & - & - &  0.53$^{+0.18}_{-0.18}$ &  0.05$^{+ 0.02}_{-0.02}$  & 0.48$^{+ 0.16}_{- 0.16}$ & 1.23$^{+0.42}_{-0.42}$ & 0.61$^{+0.21}_{-0.21}$ & 0.62$^{+0.21}_{-0.21}$ \\
\textit{Swift}$/$XRT      & 00034826016 & 1b &$<$  10.5$\times10^{3}$& 0.59$^{+ 0.29}_{-0.22}$  & 2.99 ( 16)   & 0.29$^{+0.14}_{-0.11}$  & 0.02$^{+0.01}_{-0.01}$  &  0.28$^{+0.13}_{-0.10}$ & 0.35$^{+0.16}_{-0.12}$ & 0.04$^{+0.02}_{-0.01}$  & 0.31$^{+0.14}_{-0.11}$ \\
\enddata
\tablecomments{Here, $norm_{1}$ and $norm_{2}$ are the normalization constants for two temperature components.  $\chi_{\nu}^{2}$ is the reduced $\chi^2$  and  \textit{dof} is the number of degrees of freedom. $F_{B}^{ism}$,  $F_{S}^{ism}$, and $F_{H}^{ism}$ are the ISM corrected fluxes while $F_{B}^{obs}$,  $F_{S}^{obs}$, and $F_{H}^{obs}$ are the observed X-ray fluxes of WR 125 in broad, soft, and hard energy bands, respectively.  Errors quoted on different parameters refer to the 90\% confidence level.  \\ 
$^\dagger$ The X-ray analysis tool WebPIMMS has been used to convert the average count rate given in Table \ref{tab:tab1} to fluxes in different energy bands with the specified fitting model. \\
$^\ddagger$ See Tables \ref{tab:tab2} and \ref{tab:tab3}.}
\end{deluxetable*}
\end{longrotatetable}

\begin{deluxetable}{l c c c}
\tablenum{5}
\tablecaption{NIR observations of WR 125 with TIRCAM2}\label{tab:tab5}
\tablewidth{0pt}
\tablehead{
\colhead{Date}  & \colhead{J}& \colhead{H} & \colhead{K} \\
\cline{2-4}
   \colhead{}   &   \multicolumn{3}{c}{\textbf{(mag)}}   
}
\startdata
2017-10-17 & 8.95$\pm$0.09  & 8.14$\pm$0.09  & 8.07$\pm$0.11  \\
2020-10-09 & 9.16$\pm$0.12  & 8.29$\pm$0.14  & 7.59$\pm$0.14 \\
2021-03-01 & 8.90$\pm$0.14  & 8.31$\pm$0.08  & 7.50$\pm$0.12 \\
\enddata  
\end{deluxetable}

\end{document}